\documentclass[sigconf]{acmart}

\AtBeginDocument{%
  }

\setcopyright{acmlicensed}
\copyrightyear{2026}  
\acmYear{2026}
\acmDOI{XXXXXXX.XXXXXXX}  

\acmConference[CCS '26]{Proceedings of the 2026 ACM SIGSAC Conference on Computer and Communications Security}{November 15--19, 2026}{The Hague, The Netherlands}

\acmISBN{978-1-4503-XXXX-X/XXXX/XX}

\author{
  Xing Zhang\textsuperscript{1,\textdagger}, 
  Zikang Huang\textsuperscript{2,1,\textdagger}, 
  Gang Yang\textsuperscript{3,\faEnvelope}, 
  CongChong Wang\textsuperscript{1}, 
  Lu Liu\textsuperscript{4,1}, 
  Bin Yin\textsuperscript{1}, 
  Mingyi Wang\textsuperscript{1}, 
  Ziquan Zhao\textsuperscript{1}, 
  Min Li\textsuperscript{1}, 
  Zhenyu Chen\textsuperscript{1}, 
  Bo Wu\textsuperscript{3}, 
  Lingyun Ying\textsuperscript{1,\faEnvelope}
}

\affiliation{
  \institution{
    \textsuperscript{1}QI-ANXIN Technology Research Institute, Beijing, China \\
    \textsuperscript{2}Wuhan University, Wuhan, China \\
    \textsuperscript{3}Information Support Force Engineering University, Wuhan, China \\
    \textsuperscript{4}Shandong University, Jinan, China \\[6pt]
    \textsuperscript{\textdagger}Both authors contributed equally to this research. \\
    \textsuperscript{\faEnvelope}\textit{Corresponding authors: yanggang11@nudt.edu.cn,  yinglingyun@qianxin.com}
  }
  \country{}
}


\usepackage{fontawesome}
\usepackage{graphicx}
\usepackage{amsmath}
\usepackage{algorithm}
\usepackage{algorithmicx}
\usepackage{algpseudocode}
\usepackage{cuted}
\usepackage{amsfonts}      

\usepackage{amssymb}       
\usepackage{bbold}         
\usepackage{mathrsfs}      

\usepackage{marvosym}      
\usepackage{pifont}        

\usepackage{listings}
\usepackage{fancyvrb}
\usepackage{multirow}   
\usepackage{bigstrut}   
\usepackage{makecell}
\usepackage{longtable}
\usepackage{xspace}
\usepackage{pifont}
\usepackage{xcolor} 
\usepackage{booktabs} 
\usepackage[title]{appendix}
\usepackage{graphicx}
\usepackage{seqsplit}
\usepackage[table]{xcolor}
\usepackage{pdflscape}
\usepackage{wasysym} 
\usepackage{bigfoot}
\usepackage{tcolorbox}
\usepackage{booktabs}  

\usepackage{fancyvrb}
\usepackage{array}
\usepackage{xparse}

\usepackage{array, booktabs}

\ExplSyntaxOn
\NewDocumentCommand{\splitcells}{m}
 {
  \seq_set_split:Nnn \l_tmpa_seq { | } { #1 }
  \seq_use:Nn \l_tmpa_seq { & }
 }
\ExplSyntaxOff

\hypersetup{
  colorlinks=true,
  urlcolor=ACMDarkBlue,
  linkcolor=ACMPurple,
  citecolor=ACMPurple
}

\usepackage{xcolor}
\usepackage{listings}

\usepackage{soul}  

\definecolor{modAdd}{RGB}{0,128,0}     
\definecolor{modDel}{RGB}{255,0,0}     
\definecolor{modChange}{RGB}{0,0,255}  

\newcommand{\edit}[4]{%
    \ifthenelse{\equal{#2}{add}}%
        {\textcolor{modAdd}{#3}\marginpar{\scriptsize\textcolor{modAdd}{#4:add}}}%
    {\ifthenelse{\equal{#2}{del}}%
        {\textcolor{modDel}{\cancel{#1}}\marginpar{\scriptsize\textcolor{modDel}{#4:del}}}%
        {\textcolor{gray}{\sout{#1}}\textcolor{modChange}{#3}\marginpar{\scriptsize\textcolor{modChange}{#4:mod}}}}
}

\lstdefinestyle{ultracompact}{
    language=C,
    basicstyle=\ttfamily\scriptsize,        
    backgroundcolor=\color{white},          
    commentstyle=\color{red},           
    keywordstyle=\color{blue!80!black},     
    numberstyle=\tiny\color{gray!60},       
    stringstyle=\color{green!50!black},     
    breaklines=true,
    captionpos=b,
    numbers=left,
    numbersep=4pt,                          
    showspaces=false,
    showstringspaces=false,
    showtabs=false,
    tabsize=2,
    frame=none,
    xleftmargin=0pt,
    xrightmargin=0pt,
    belowskip=1pt,
    lineskip=1pt                            
}

\usepackage{xurl}
\usepackage[hyphenbreaks]{breakurl}

\newcommand{\tool}{\textsc{SynapseFlow}\xspace}
\renewcommand{\shortauthors}{Xing Zhang, Zikang Huang, Gang Yang, Lingyun Ying, and et al.}

\begin{document}
%


\title[SYNAPSEFlow]{Thinking More, Harnessing Better: Automatic Harness Generation with Dataflow Aggregation and Workflow Decomposition}

\begin{abstract}

High-quality fuzz harnesses are essential for effective gray-box fuzzing. 
While Large Language Models (LLMs) offer promise for automating this task, existing one-turn generation methods suffer from hallucinations and inadequate coverage due to coarse-grained function targeting and misaligned generation workflows. 
We present \tool, an automatic harness generator that addresses these limitations through two key innovations: dataflow-aware function aggregation and a staged, rollback-enabled generation workflow decomposition. 
\tool first analyzes source code to construct Structural Flow Graphs and extract coherent Function Triplets. 
It then synthesizes harnesses via a decomposed four-stage process governed by a staged rollback algorithm to ensure correctness. 
We evaluated \tool on 25 real-world open-source software projects.
The experimental results indicate that \tool outperforms state-of-the-art tools (OSS-Fuzz-Gen, CKGFuzzer, PromeFuzz), achieving 3.07$\times$, 1.71$\times$, and 4.26$\times$ higher branch coverage, and 1.77$\times$, 1.51$\times$, and 1.36$\times$ higher bug detection rates, respectively. 
Most importantly, \tool discovered 7 previously unreported bugs (5 assigned CVEs), demonstrating its practical effectiveness in real-world bug discovery.

\end{abstract}

\maketitle

%

\lstset{
	xleftmargin=1em,
	xrightmargin=1em,
	captionpos=b,
	breaklines=true,
	caption = {this is caption},
}
\section{Introduction}
Fuzzing is a cornerstone technique in modern software security analysis.
Among its various paradigms, white-box fuzzing, which leverages source code analysis to guide test generation, offers the potential for deep program exploration.
High-quality fuzz harnesses—code segments that invoke target functions with fuzzing inputs—are pivotal for effective function-oriented gray-box fuzzing.
They enable direct testing of internal program logic, circumventing the coverage limitations often encountered when fuzzing only through external application interfaces.

This need is particularly urgent for C projects, which constitute the backbone of critical system software, libraries, and embedded systems. Due to the prevalence and severity of memory safety vulnerabilities in C, achieving comprehensive coverage is essential~\cite{sokEternalWarinMemory}. 
However, modern C applications and libraries exhibit intricate function dependencies, making comprehensive coverage via whole-application fuzzing challenging~\cite{manes2019art}. 
Consequently, function-oriented fuzzers, which rely on specialized harnesses to test individual functions directly, have gained widespread adoption~\cite{serebryany2016continuous}.


The core challenge lies in automatically synthesizing such harnesses, which demands a deep understanding of source code to logically compose relevant functions. Large Language Models (LLMs), with their strong code comprehension and generation capabilities, have thus emerged as promising candidates for this task~\cite{fakhoury2024llm, Liu_OSS-Fuzz-Gen_Automated_Fuzz_2024}.
Prior research endeavors have integrated prompt engineering with proprietary LLMs to generate harnesses through different combinations of API mutations~\cite{zhang2023understanding}.
Those methods employ zero-shot or few-shot prompting and have also explored various code composition strategies~\cite{zhang2024effective,zhang2023understanding,lyu2024prompt,xu2024ckgfuzzer}.
However, prevalent LLM-based approaches are hampered by a monolithic generation paradigm, leading to two core shortcomings: 1)   imprecise, often ad-hoc function selection and 2) severe LLM hallucinations.
Firstly, lacking systematic dataflow analysis leads to ad-hoc function selection (e.g., misidentifying internal functions), which fails to form a minimal, coherent set for effective testing. 
This yields harnesses that are either incomplete or burdened with irrelevant code.
Secondly, the monolithic generation task commonly overwhelms LLMs, causing hallucinations  (e.g., generating incorrect or redundant code) that cripple the harness’s correctness and effectiveness.  
These limitations collectively result in low code coverage and missed vulnerabilities.

Therefore, generating high-quality harnesses presents a dual challenge:
1) intelligently identifying and aggregating a functionally coherent cluster of functions, and 2) orchestrating a robust, multi-step synthesis process that mitigates LLM errors.
This process inherently involves a trade-off between granularity (testing focused units) and comprehensiveness (covering enough logic to trigger deep bugs)~\cite{gorz2025empirical}.
More importantly, a well-designed harness should ensure that any crash it produces during fuzzing is a meaningful, reproducible bug in the original application context, not an artifact of the harness itself.
Thus, automated harness generation can be viewed as a constrained optimization and planning problem over the project's function call graph.

To address these issues, we propose a paradigm shift: moving from monolithic LLM invocation to a structured, decomposition-based workflow akin to expert human reasoning.
This workflow first \textit{digests} the project to form semantically meaningful function groups, then \textit{decomposes} the code writing into manageable, verifiable steps to synthesize a high-coverage harness.
Furthermore, the process incorporates feedback and rollback mechanisms inspired by task decomposition for AI agents~\cite{huang2024understanding} to contain errors and hallucinations.

Following this rationale, achieving high-quality, automated gray-box harness generation necessitates addressing the following two core research questions:

\textbf{How to identify and group involved functions for harness generation from source code?}
This requires analyzing source code to identify functions that process security-critical data (often from external inputs) and group them logically.
A key challenge is to move beyond simple metrics (e.g., complexity) and understand dataflow, parameter semantics, and functional cohesion to form minimal yet sufficient groups.
Current LLM-only approaches lack the precise program analysis for this.

\textbf{How to mitigate the risks of hallucinations and error accumulation in the long-chain generation workflow?}
When tasked with generating complex harness code in one turn, LLMs frequently hallucinate details: they may invent non-existent functions, misuse macros, or insert inefficient I/O operations (e.g., \texttt{fprintf}), leading to compilation failures or crippled fuzzing performance.
While some methods inject knowledge via prompts~\cite{xu2024ckgfuzzer, promefuzz}, this proves inadequate for long, complex generation chains.
A robust solution requires decomposing the workflow into sequential, verifiable sub-tasks with built-in error detection and recovery.

To tackle these challenges, we propose \tool, an LLM-powered framework grounded in \textit{dataflow aggregation} and \textit{workflow decomposition} for automatic harness generation.
It first employs hybrid static analysis and LLM reasoning to digest the whole project and construct minimal, coherent function groups. 
It then orchestrates a multi-stage generation workflow with a staged rollback algorithm to ensure correctness and mitigate hallucinations.
This constitutes a paradigm shift from prior monolithic generation: generation steps are dictated by extracted dataflow, and the workflow self-corrects via targeted rollback to isolate hallucinations.
We evaluate \tool on 25 real-world, open-source C projects.
It significantly outperforms state-of-the-art (SOTA) tools (OSS-Fuzz-Gen~\cite{Liu_OSS-Fuzz-Gen_Automated_Fuzz_2024},  CKGFuzzer~\cite{xu2024ckgfuzzer}, and PromeFuzz~\cite{promefuzz}), achieving 3.07x, 1.71x, and 4.26x higher branch coverage and 1.77x, 1.51x, and 1.36x higher crash discovery rates, respectively. 
Notably, even on these extensively fuzzed projects, harnesses generated by \tool led to the discovery of 7 previously unreported bugs (4 assigned CVEs), demonstrating its ability to find deep, overlooked Bugs.

This paper makes the following contributions:
\begin{itemize}
    \item \textbf{A Method for Dataflow-Aware Function Aggregation}: We propose a novel method integrating lightweight static dataflow analysis with LLM-based semantic reasoning to identify and aggregate functionally related functions into minimal, self-contained groups. This provides a sound foundation for harness generation beyond API-level testing.

    \item \textbf{A Framework for Robust, Decomposed Synthesis}: We design a stepwise and reversible prompting framework governed by a staged rollback algorithm. This decomposes complex harness synthesis into a sequence of manageable LLM sub-tasks, effectively mitigating hallucinations and error propagation common in single-prompt approaches.

    \item \textbf{Implementation and Extensive Evaluation}: We implement our approach in \tool, an automated harness generation tool for C programs. Extensive evaluation on real-world projects shows that \tool-generated harnesses achieve higher coverage and uncover more bugs than SOTA tools, including the discovery of 7 previously unreported bugs.
\end{itemize}
\section{Background and Motivation}

\lstdefinestyle{mycstyle}{
	xleftmargin=1em,
	xrightmargin=1em,
	captionpos=b,
	breaklines=true,
        mathescape=true,
        keywordstyle=\color{red},
        keywordstyle=[2]\color{blue},
        morekeywords={j40_from_memory},
        morekeywords=[2]{j40_output_format, j40_next_frame, j40_current_frame, j40_frame_pixels, j40_row},
    basicstyle=\ttfamily\tiny,
	caption = {this is caption}
}


\subsection{Harness in Gray-Box Fuzzing}
\label{subsec:h}
In the context of gray-box fuzzing for C source projects, a fuzz harness (or fuzz driver) is a specialized piece of code that acts as a test interface for a specific functional unit.
Its primary role is to receive raw fuzzing input (typically a byte array), transform this input into the appropriate program state (e.g., by initializing data structures), and then invoke a sequence of target functions to test their behavior under varied inputs.
By isolating and directly testing internal components, harnesses enable more focused and efficient vulnerability discovery compared to whole-application fuzzing.

A high-quality harness must satisfy several critical criteria to be effective in practice:
\begin{itemize}
    \item \textbf{Functional Completeness:} It must invoke a logically coherent and \textit{minimal} set of functions that together implement a meaningful feature. Omitting necessary functions leads to low coverage; including irrelevant ones adds noise and reduces fuzzing efficiency.
    \item \textbf{Semantic Correctness:} The harness must respect the usage specifications and data dependencies of each invoked function (e.g., correct parameter types, proper initialization order, and valid state transitions).
    \item \textbf{Syntactic Soundness:} The generated code must be compilable and free of syntax errors. It should also avoid introducing constructs that hinder fuzzing performance, such as file I/O operations or excessive logging.
\end{itemize}
Crafting such a harness manually requires deep understanding of the codebase and is labor-intensive, motivating the need for reliable automation. 
Existing automated methods, as we discuss next, often fall short of meeting these quality standards.

\subsection{Harness Generation}
Automated harness generation techniques can be broadly categorized into two paradigms: traditional program-analysis-based methods and modern LLM-based approaches.

\noindent{\textbf{Traditional methods}} (e.g., template-based~\cite{2021IntelliGen}, slicing-based~\cite{FUDGE,2020FuzzGen}) rely on static/dynamic analysis. 
Their synthesis methods are deterministic but rigid, often failing to generate complex logic and lacking semantic understanding, which leads to poor \textit{functional completeness}.

\noindent{\textbf{LLM-based methods}} (e.g., OSS-Fuzz-Gen, CKGFuzzer, PromeFuzz) leverage code comprehension to produce more sophisticated code. 
However, they introduce new critical flaws: 1) their function selection strategies are ad-hoc, lacking principled, dataflow-aware aggregation of functions; 2) they predominantly use a monolithic, one-shot generation workflow, which is prone to LLM hallucinations and error cascades without recovery mechanisms. 
Consequently, they often compromise semantic correctness and functional completeness.

In essence, existing methods fall short because they fail to address two core aspects of the problem: systematic function grouping and robust, error-tolerant generation.

\subsection{Motivation Example}
To ground the aforementioned challenges in a concrete scenario, we use the \texttt{j40} project~\cite{url_j40}, a widely-used C library for parsing JPEG XL images, as a running example.
Constructing an effective harness for \texttt{j40} requires invoking a specific sequence of functions that manipulate core data structures, as shown in Figure~\ref{fig:three_listings_direct} (Listing~\ref{example_j40}) and as listed in Table~\ref{j40_api}.

\begin{figure*}[!t]
    \centering
    \begin{minipage}[t]{0.32\textwidth}
        \centering

\begin{lstlisting}[style=mycstyle, caption={An idealical example harness}, label=example_j40]
int LLVMFuzzerTestOneInput(const uint8_t *data, size_t size) {
  j40_image image;
  j40_from_memory(&image,(void*)data,size,NULL);
  j40_output_format(&image, J40_RGBA, J40_U8X4);
  while(j40_next_frame(&image)){	
    j40_frame frame = j40_current_frame(&image);
    j40_pixels pixels;
    pixels = j40_frame_pixels(&frame, J40_RGBA);
    for (int y = 0; y < pixels.height; ++y) {
        j40_row(pixels, y);}}
  j40_free(&image);
  return 0;}
\end{lstlisting}

\begin{lstlisting}[style=mycstyle, caption={A harness generated by OSS-Fuzz-Gen}, label=example_j40_oss]
int LLVMFuzzerTestOneInput(const uint8_t *data, size_t size) {
    j40_image *image;
    j40_frame frame;
    if (size < sizeof(j40_image)) {
        return 0;}
    image = (j40_image *)data;
    frame = j40_next_frame(image);
    return 0;}
\end{lstlisting}

    \end{minipage}
    \hfill
    \begin{minipage}[t]{0.32\textwidth}
        \centering

\begin{lstlisting}[style=mycstyle, caption={A harness generated by CKGFuzzer}, label=example_j40_ckg]
int LLVMFuzzerTestOneInput(const uint8_t *data, size_t size) {
    j40_image image;
    j40_err err;
    if (size < 1) {
        return 0;}
    void* input_buffer = malloc(size);
    if (!input_buffer) {
        return 0;}
    memcpy(input_buffer, data, size);
    memset(&image, 0, sizeof(image));
    err = j40_from_memory(&image, input_buffer, size, nullptr);
    if (err == 0) { 
        int result = j40_next_frame(&image);}
    free(input_buffer);
    return 0;}
\end{lstlisting}

    \end{minipage}
    \hfill
    \begin{minipage}[t]{0.32\textwidth}
        \centering

\begin{lstlisting}[style=mycstyle, caption={A harness generated by PromeFuzz}, label=example_j40_prom]
int LLVMFuzzerTestOneInput(const uint8_t *Data, size_t Size) {
    j40_image* image = (j40_image*)calloc(1, sizeof(j40_image));
    if (!image) return 0;
    if (Size > 0) {
        j40_err err = j40_from_memory(image, (void*)Data, Size, NULL);
        if (err != 0) {
            const char* err_str = j40_error_string(image);} 
        else {
            while (j40_next_frame(image) == 0) {}}}
    FILE* input = fopen("input", "wb");
    if (input) {
        fwrite(Data, 1, Size, input);
        fclose(input);
        j40_err e = j40_from_file(image, "input");
        if (e != 0) {
            const char* err_str = j40_error_string(image);}}
    j40_free(image);
    return 0;}
\end{lstlisting}

    \end{minipage}
    \caption{A comparison of harnesses generated by distinct approaches for \texttt{j40} project.}
    \label{fig:three_listings_direct}
\end{figure*}

\begin{table}[h]
    \centering
\caption{Functions with key data structures in the \texttt{j40} project.}
\label{j40_api}
    \begin{tabular}{clll}
\toprule
         \textbf{Seq.}&\textbf{Function Name}&  \textbf{Input Struct}& \textbf{Output Struct}\\
    \hline
    \rowcolor[HTML]{EFEFEF}
         1 & \texttt{j40\_from\_memory} &  \texttt{ByteStream}& \texttt{j40\_image}\\
         2 & \texttt{j40\_output\_format}&  \texttt{j40\_image}& \\
 \rowcolor[HTML]{EFEFEF}
         3 & \texttt{j40\_next\_frame}&  \texttt{j40\_image}& \\
         4 & \texttt{j40\_current\_frame}&  \texttt{j40\_image}& \texttt{j40\_frame}\\
\rowcolor[HTML]{EFEFEF}
         5 & \texttt{j40\_frame\_pixels}&  \texttt{j40\_frame}& \texttt{j40\_pixels}\\
         6 & \texttt{j40\_row}&  \texttt{j40\_pixels}& \\
\rowcolor[HTML]{EFEFEF}
         7 & \texttt{j40\_free}&  \texttt{j40\_image}& \\
\bottomrule
    \end{tabular}
\end{table}

Figure~\ref{fig:three_listings_direct} contrasts a high-quality, developer-written harness (Listing \ref{example_j40}) with harnesses generated by SOTA LLM-based tools (Listings \ref{example_j40_oss}, \ref{example_j40_ckg}, \ref{example_j40_prom}).
The ideal harness (Listing \ref{example_j40}) demonstrates \textit{functional completeness} by correctly sequencing all seven key functions, following the dataflow from \texttt{j40\_image} to \texttt{j40\_pixels}.
It also exhibits \textit{semantic correctness} (e.g., proper initialization before use, a loop to process multiple frames) and \textit{syntactic soundness} (no extraneous operations).

In contrast, the auto-generated harnesses reveal the limitations of current methods:
\begin{itemize}
    \item OSS-Fuzz-Gen (Listing~\ref{example_j40_oss}) focuses on single-function generation. It completely misses the necessary dataflow, selecting only two unrelated functions and producing a harness that cannot meaningfully test the library (line 7), failing \textit{functional completeness}.
    \item CKGFuzzer (Listing~\ref{example_j40_ckg}) attempts to use external knowledge but lacks fine-grained dataflow understanding. It generates a plausible but incomplete sequence (only two functions in line 11, 13) and introduces unnecessary operations like manual memory allocation/copying (\texttt{malloc}, \texttt{memcpy}), which is noisy and potentially incorrect, thus compromising \textit{semantic correctness}.
    \item PromeFuzz (Listing~\ref{example_j40_prom}) identifies more related functions (line 7, 17) yet still fails to cover the full group. It erroneously includes \texttt{j40\_from\_memory} and \texttt{j40\_from\_file} (line 5, 14), introducing two external input handlers and violating the single-entry principle of libfuzzer. The added file operations (line 10) further compromise fuzzing efficiency.
\end{itemize}

\begin{figure}[htp]
	\centering
	\includegraphics[scale=0.8]{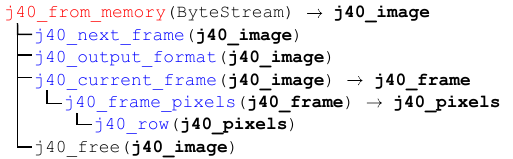}
    \caption{Structural flow graph for j40 depicting data structure transformations between key APIs, which underpins the logical grouping of functions for harness generation.}
	\label{example_sf_j40}
\end{figure}

\noindent The core issue is that these tools fail to identify the relationships formed by input and output structures between functions. We define the \textit{structural flow graph (SFG)} as the representation of these structural dependencies that logically connects functions via their struct parameters. Without this perspective, existing tools cannot derive the SFG (illustrated in Figure~\ref{example_sf_j40}) required to group the seven functions correctly.
Furthermore, their monolithic generation workflows offer no chance to detect and rectify the introduced hallucinations or logical errors.

\subsection{Problem Definition}
\label{sec:problem-definition}

The motivating example crystallizes the core task of this work: automatically generating high-quality fuzz harnesses for C source code projects. 
To achieve this automatically, the process must fundamentally address two intertwined sub-problems:

\noindent{\bf{Coherent Function Grouping:}} Given a source project \(S\), identify a set of functions \(G = \{f_1, f_2, ..., f_k\}\) that collectively implement a distinct, testable feature. 
The grouping must be grounded in the actual \textit{dataflow} between functions (via shared data structures), not just call relations or semantic similarity, to ensure functional completeness.

\noindent{\bf{Robust Harness Synthesis:}} Given a well-formed function group \(G\), synthesize a harness \(H\) that correctly sequences the calls to functions in \(G\), manages all necessary data structures, and adheres to C semantics. 
The synthesis process must be robust against LLM hallucinations and errors.



\noindent To tackle the grouping issue, we require a principled model for understanding and categorizing function-data interactions. 
To group functions effectively based on dataflow, we analyze their roles in processing data structures. 
We define three semantic categories for functions in a project:

\begin{itemize}
    \item \textbf{Input Stream Function (ISF)}: A function that serves as a unique entry point, consuming unstructured, external data (e.g., from memory, a file) and producing or initializing an internal data structure, e.g., \texttt{j40\_from\_memory} in the motivation example.
    \item \textbf{Helper Function (HPF)}: A function responsible for the lifecycle management (allocation, initialization, deallocation) of a data structure, e.g., \texttt{j40\_free} and \texttt{j40\_from\_memory} in the motivation example.
    \item \textbf{Process Function (PRF)}: A function that transforms, validates, or reads the content of an already-initialized data structure but does not manage its lifecycle, e.g., \texttt{j40\_next\_} \texttt{frame}, \texttt{j40\_frame\_pixels} and \texttt{j40\_}\texttt{row} in the motivation example.
\end{itemize}

\noindent  Noteworthy, a function can belong to multiple categories simultaneously, e.g., \texttt{j40\_from\_memory} in the motivation example is both an ISF and an HPF.

Building upon these categories, we define \textit{Function Triplet (FT)} as the atomic unit for harness generation. 
An FT represents a minimal, self-contained data processing unit centered around one primary data structure flow.
Formally, an FT is an ordered triplet:
\[
FT = (I, P, H)
\]
subject to the following constraints:
\begin{itemize}
\item \(I\) is a unique singleton \textit{ISF}.  
\item \(P\) is an optional set of \textit{PRF}s.  
\item \(H\) is an optional set of \textit{HPF}s.  
\end{itemize}

\noindent We can formalize FT as:
\[
FT = (I, P, H) \quad \textbf{where} \quad
\begin{cases}
|I| = 1 \\
I \subseteq \mathcal{F}_{ISF} \\
P \subseteq \mathcal{F}_{PRF} \cup \emptyset \\
H \subseteq \mathcal{F}_{HPF} \cup \emptyset
\end{cases}
\]
where \(\mathcal{F}_{ISF}\), \(\mathcal{F}_{PRF}\) and \(\mathcal{F}_{HPF}\) denote the sets of all ISFs, PRFs and HPFs in the project, respectively, and \(|\cdot|\) denotes set cardinality.

A source project \(S\) can be digested into multiple, distinct FTs, each anchored by a different ISF:
\[
S \rightarrow \{ FT_1, FT_2, \dots, FT_n \},
\]
where each \(FT_i = (I_i, P_i, H_i)\) and \(\forall i \neq j, I_i \neq I_j\).

The process of discovering these FTs from source code directly addresses the grouping issue. 
The FT provides a semantically coherent function group (\(G = I \cup P \cup H\)) that is grounded in dataflow (the input/output structures connecting \(I\), \(P\), and \(H\)). 
For our running example, the functions in Table~\ref{j40_api} naturally form a single FT: \(I=\{\texttt{j40\_from\_memory}\}\), while \(P = \{\texttt{j40\_output\_format},\allowbreak \texttt{j40\_next\_frame}, \allowbreak \texttt{j40\_current\_frame},\allowbreak \texttt{j40\_frame\_pixels},\allowbreak \texttt{j40\_}\)

\noindent \(\texttt{row}\}\), \(H=\{\allowbreak \texttt{j40\_from\_memory}, \allowbreak \texttt{j40\_free}\}\). 
This model elegantly captures the functional completeness requirement.

\subsection{Challenges}
\label{subsec:chall}
Given the problem definition above, realizing an automated solution like \tool involves overcoming two major implementation challenges that correspond to sub-problems (in Section~\ref{sec:problem-definition}):

\noindent{\bf{Challenge 1: Dataflow-Aware Function Grouping.}}
How can we accurately identify ISF, HPF, and PRF functions from C source code and cluster them into correct FTs? 
This requires going beyond syntactic call graphs to understand how data structures are created, transformed, and passed between functions—a task that demands a fusion of lightweight static analysis for structural tracking and LLM-based reasoning for semantic role disambiguation.

\noindent{\bf{Challenge 2: Hallucination-Robust Multi-Stage Synthesis.}}
Given an FT, how can we design a generation workflow that decomposes the complex harness synthesis task into a sequence of simpler, verifiable sub-tasks (e.g., structure definition, initialization, function chaining, cleanup)? 
This workflow must incorporate mechanisms for error detection and recovery (e.g., rollback) to prevent hallucinations and error cascades, ensuring the final harness is both semantically correct and syntactically sound.

The design of \tool, presented in the next section, is our concrete response to these challenges.
\section{Methodology}
\label{sec:meth}
\subsection{Overview}
\label{subsec:overview}

\begin{figure*}[htbp]
	\centering
    \includegraphics[width=1.00\linewidth]{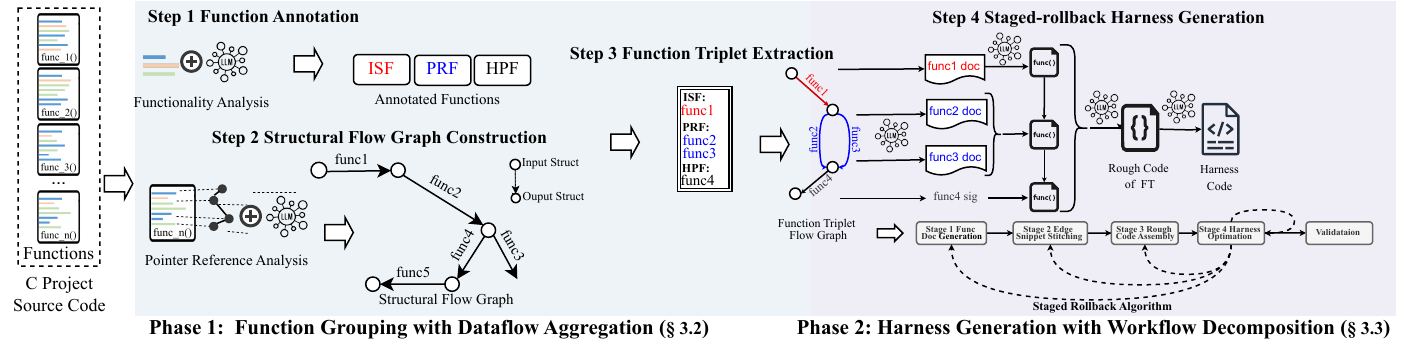}
	\caption{The workflow of \tool.}
	\label{overview}
\end{figure*}

\tool automates high-quality harness generation for source code by decomposing the process into two core phases (Figure~\ref{overview}) that directly address the challenges defined in Section~\ref{subsec:chall}.

\noindent{\bf{Phase 1: Function Grouping with Dataflow Aggregation.}} 
This phase digests the source project (Section~\ref{subsec:fa}) to extract coherent FTs—minimal, testable function groups grounded in dataflow. 
The process involves classifying functions by semantic role (ISF/PRF/HPF), constructing a SFG to model structure propagation (Section~\ref{subsec:sfg}), and applying graph algorithms to extract FTs (Section~\ref{subsec:fe}).

\noindent{\bf{Phase 2: Harness Generation with Workflow Decomposition.}}
Given an FT, this phase synthesizes the final harness. 
To mitigate the hallucinations and errors inherent in monolithic LLM prompting, we decompose synthesis into a sequence of four simpler, verifiable stages (Section~\ref{subsec:stage-design}).
Meanwhile, a \textit{staged rollback algorithm} (Section~\ref{subsec:rollback}) orchestrates this workflow, enabling recovery from failures by rolling back to previous stages.

\noindent{\bf{Running Example.}}
We continue with the \texttt{j40} project. Phase 1 constructs its SFG (Figure~\ref{example_sf_j40}) and extracts an FT (Table~\ref{j40_tfg}) containing the seven key API functions. 
Phase 2 then transforms this FT into the high-quality harness of Listing~\ref{example_j40}. 
The following subsections detail each phase.

\begin{table}[h]
\centering
\caption{Function Triplet of the \texttt{j40} project. Functions are annotated with their types:  \textcolor{red}{ISF},  \textcolor{blue}{PRF}, and HPF.}
\label{j40_tfg}
\scalebox{1}{
\begin{tabular}{lll}
\toprule
\textbf{Function Name} & \textbf{Type} & \textbf{Structural Flow}  \\
\hline
\rowcolor[HTML]{EFEFEF}
\textcolor{red}{\texttt{j40\_from\_memory}}   & \textcolor{red}{ISF}\&{HPF}  & (null), \texttt{j40\_image} \\
\textcolor{blue}{\texttt{j40\_output\_format}} & \textcolor{blue}{PRF}  & \multirow{2}{*}{\texttt{j40\_image}, (null)} \\
\textcolor{blue}{\texttt{j40\_next\_frame}}    & \textcolor{blue}{PRF}  &   \\
\rowcolor[HTML]{EFEFEF}
\textcolor{blue}{\texttt{j40\_current\_frame}} & \textcolor{blue}{PRF}  & \texttt{j40\_image}, \texttt{j40\_frame}  \\
\textcolor{blue}{\texttt{j40\_frame\_pixels}}  & \textcolor{blue}{PRF}  & \texttt{j40\_frame}, \texttt{j40\_pixels}  \\
\rowcolor[HTML]{EFEFEF}
\textcolor{blue}{\texttt{j40\_row} }           & \textcolor{blue}{PRF}  & \texttt{j40\_pixels}, (null)  \\
\texttt{j40\_free} & HPF &  \\
\bottomrule
\end{tabular}}
\end{table}

\subsection{Function Triplet Extraction with Dataflow Aggregation}

This phase aims to automatically discover FTs from the source code.
An FT represents a minimal, coherent set of functions that should be tested together, grounded in their dataflow dependencies.
The extraction involves three main steps: 1) annotating each function with its semantic type (ISF, PRF, or HPF), 2) constructing an SFG that models data structure propagation between functions, and 3) applying a graph algorithm to the SFG to extract the FTs.
We now elaborate on each step.

\subsubsection{Function Annotation}

\label{subsec:fa}
The goal of this step is to label each function in the project as an ISF, PRF, or HPF according to the definitions in Section~\ref{subsec:overview}.
Accurate annotation requires understanding both syntactic features (parameter types) and functional semantics (what the function does).
We therefore employ a hybrid approach that combines lightweight static analysis for syntax with LLM-based reasoning for semantics.

\noindent{\bf Identifying ISFs.}
We first use syntax analysis to filter functions that have pointer-type parameters (e.g., \texttt{void*}, \texttt{uint8\_t*}) which could represent raw byte streams.
For these candidate functions, we need to distinguish true byte-stream handlers from those that process semantic strings (e.g., filenames).
We leverage the LLM's semantic understanding through a set of distinct prompt templates (detailed in Appendix~\ref{subsec:paraanal}).
These prompts ask the LLM to judge, in different formats, whether a specific parameter represents a contiguous byte input.
Next, an LLM uses multiple distinct prompts (detailed in
Appendix~\ref{subsec:inparadeter}) to filter functions that reference
specific meaningful strings, such as those passed by references, filenames, or path names.
A voting mechanism on the LLM's responses yields the final identification of ISFs.

\noindent{\bf Distinguishing PRFs and HPFs.}
This is a two-stage process.
First, syntax analysis isolates functions that operate on structs (have struct parameters or return a struct).
Second, for these candidate functions, specialized LLM prompts (Appendix~\ref{subsec:funcanal}) infer their functional role.
Functions described as performing initialization or cleanup of a struct are labeled HPFs.
Functions described as manipulating, transforming, or reading an already-initialized struct are labeled PRFs.
A single function can possess multiple labels (e.g., a function that reads a stream and allocates a struct is both an ISF and an HPF).
We treat these labels as descriptive attributes.

\subsubsection{Structural Flow Graph Construction}

\label{subsec:sfg}
The goal of this step is to construct an SFG, a directed graph that provides a high-level abstraction of how data structures flow between functions.
This graph is crucial for understanding the connectivity and dependencies that will guide FT extraction.

\noindent{\bf Structure Directionality Inference.}
To build the SFG, we must first determine each function's input and output structures.
Return types and non-pointer struct parameters are straightforward.
For ambiguous pointer-to-struct parameters, we need to infer if they are input (read-only), output (write-only), or both.
We resolve this by prompting the LLM to analyze the function's operational semantics on that parameter.

\noindent{\bf Graph Formalization.}
With input/output structures identified, we formalize the SFG as a directed graph \(G = (V, E)\):
\begin{itemize}
    \item \textbf{Nodes (\(V\)):} Each node represents a unique structure type. A special ``(null)'' node represents the absence of an input/output structure.
    \item \textbf{Edges (\(E\)):} A directed edge \(e = \langle n_{in}, n_{out} \rangle \in E\) represents a function \(f\) that consumes an input structure \(n_{in}\) and produces an output structure \(n_{out}\). The edge is labeled with \(f\).
\end{itemize}
By processing all project functions, we construct the complete SFG.
For our running example, the SFG for \texttt{j40} is visualized in Figure~\ref{example_sf_j40}, clearly showing the flow from \texttt{ByteStream} to \texttt{j40\_pixels}.

\subsubsection{Function Triplet Extraction}

\label{subsec:fe}
The goal of this final step is to extract one FT for each unique ISF in the project.
The ISF serves as the natural entry point and anchor for a harness.
Our extraction algorithm, detailed in Algorithm~\ref{algo:selection}, works on the annotated SFG.

\noindent{\bf Extraction Process.}
For a given ISF \(f_{ISF}\), the algorithm first prunes the SFG by removing edges corresponding to all \textit{other} ISFs (their harnesses will be generated separately).
From the pruned graph, it performs a forward dataflow analysis starting from the ISF's output structure, collecting all reachable functions.
It also performs a backward analysis if needed.
The functions within this reachable subgraph are then filtered based on their annotations: all PRFs are included, and relevant HPFs (e.g., destructors for allocated structures) are added.
The resulting set \(\{f_{ISF}\} \cup P \cup H\) forms the FT.
A function appearing in multiple FTs (e.g., a common \texttt{free} function) is handled appropriately in each harness.

\noindent{\bf Handling Multi-role Functions.}
A key nuance is handling functions with multiple annotations (e.g., a function that is both an ISF and a PRF).
During pruning, only its ISF role is removed from the graph for other FTs; it remains present for its PRF role.
If a function in the final FT is both a PRF and an HPF, it is treated as a PRF for generation priority, ensuring the core processing logic is captured.

\begin{algorithm}
    \caption{Function Triplet Extraction Algorithm}
    \label{algo:selection}
    \begin{algorithmic}[1]
        \Require \(f_{ISF}=(n_{in}, n_{out})\), \(\mathcal{F}_{ISF}\),\(\mathcal{F}_{PRF}\),\(\mathcal{F}_{HPF}\), Structural Flow Graph \(G\).
        \Ensure \textit{FT} for \(f_{ISF}\).
        \State \(newG \leftarrow G\).remove\_edges\_from(\(\mathcal{F}_{ISF} - \{f_{ISF}\}\))
        \State \(InNodes  \leftarrow newG\).ancestors(\(n_{in}\)) + \(n_{in}\)
        \State \(OutNodes  \leftarrow newG\).descendants(\(n_{out}\)) + \(n_{out}\)
        \State \(FTGraph \leftarrow newG\).subgraph(\(InNodes+OutNodes\))
        \State \(Edges \leftarrow FTGraph\).edges()
        \State \(F_{PRF} \leftarrow \{f| f\in Edges, f \in \mathcal{F}_{PRF}\}\)
        \State \(F_{HPF} \leftarrow \{f| f\in Edges, f \in \mathcal{F}_{HPF} ,f \notin F_{PRF} \}\)
        \State \(FT \leftarrow (f_{ISF},F_{PRF},F_{HPF})\)
        \State \Return \(FT\)
    \end{algorithmic}
\end{algorithm}

\subsection{Harness Generation with Staged Decomposition and Rollback}
This phase takes an FT as input and synthesizes the corresponding fuzz harness.
The core insight is that generating the entire harness in one LLM call is error-prone.
Therefore, we decompose the task into a sequence of four simpler stages, each with a clear objective and validation criteria.
A \textit{staged rollback algorithm} orchestrates these stages, providing fault tolerance by allowing the process to roll back to a previous stage upon failure. The detail theoretical proof of the advantages of decomposition and rollback is shown in Appendix~\ref{appendix1}.

\subsubsection{Stage Design}
\label{subsec:stage-design}

The stage decomposition follows a progressive refinement strategy.
We transform the complex goal of “generate a full harness” into simpler sub-goals: first understand the functions, then generate code for local data transformations, then assemble the pieces, and finally polish the result.
This approach aligns with the LLM's capabilities and limits context length per step.

\noindent{\bf Stage 1: Function Documentation Generation.}
The goal is to convert the raw source code of each function in the FT into structured, textual API documentation.
This includes the function signature, a description of its purpose, its usage scenario, and example invocation patterns.
This step abstracts away low-level syntax, providing the LLM with high-level semantic knowledge for subsequent code synthesis (by employing prompts in Appendix~\ref{subsec:funcdoc}).
For our \texttt{j40} example, this stage converts the seven functions in Table~\ref{j40_tfg} into their corresponding API documentation.

\begin{figure}[htp]
	\centering
	\includegraphics[scale=0.8]{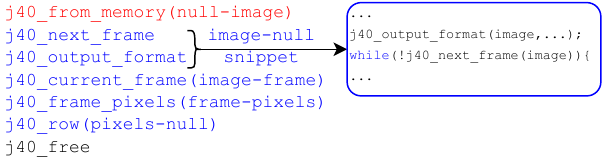}
	\caption{Structural processing snippet generation for the \texttt{j40} project (Stage 2). Each snippet corresponds to a coherent data transformation identified from the FT in Table~\ref{j40_tfg}.}
	\label{step2_j40}
\end{figure}

\noindent{\bf Stage 2: Structure Snippet Stitching.}
The goal is to generate small, correct code snippets for localized data transformations.
Using the SFG, we identify groups of functions that share the same input and output structure (forming a processing ``unit'').
For each unit, the LLM is prompted to generate a code snippet that correctly sequences the involved function calls to transform the input structure into the output structure.
This breaks down the global code generation problem into manageable local problems.
Figure~\ref{step2_j40} illustrates this process for the five snippets identified in \texttt{j40}'s FT (Table~\ref{j40_tfg}).
The prompt used is listed in Appendix~\ref{subsec:stitch}.

\begin{figure}[htp]
	\centering
	\includegraphics[scale=0.81]{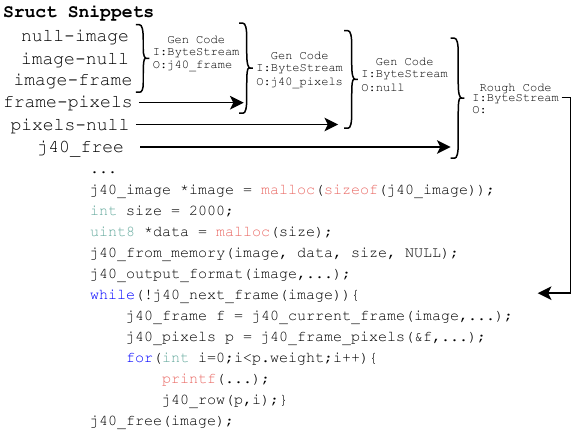}
	\caption{Rough code assembly for the \texttt{j40} project (Stage 3). Structural snippets are iteratively merged following the dataflow order.}
	\label{step3_j40}
\end{figure}

\noindent{\bf Stage 3: Rough Code Assembly.}
The goal is to combine the individual snippets from Stage 2 into a complete, executable prototype.
The LLM iteratively merges adjacent snippets (following the dataflow order in the SFG) by writing code that connects their inputs and outputs using the prompt in Appendix~\ref{subsec:codeass}.
The output of this stage is a rough but complete code sequence that correctly invokes all functions in the FT.
Figure~\ref{step3_j40} shows the assembly process for \texttt{j40}, where the five snippets are logically merged. 


\noindent{\bf Stage 4: Code Optimization and Harness Transformation.}
The goal is to convert the rough prototype into a final, fuzzing-ready harness.
The LLM performs several tasks: it wraps the code into the standard \texttt{LLVMFuzzerTestOneInput} function signature, removes any operations harmful to fuzzing performance (e.g., file I/O, excessive prints), and ensures proper resource cleanup.
The output is the final harness code, which for \texttt{j40} matches the high-quality harness shown in Listing~\ref{example_j40}.

\subsubsection{Staged Rollback Algorithm}
\label{subsec:rollback}
Decomposing the workflow improves quality but introduces more steps.
To manage potential failures efficiently and avoid restarting from scratch, we introduce the staged rollback algorithm (Algorithm~\ref{algo:generation}).
Its core idea is simple: if the final harness (Stage 4 output) fails compilation or a basic runtime test, instead of discarding all previous work, we roll back to the output of Stage 3 and regenerate Stage 4.
If repeated failures occur, we roll back further (to Stage 2, then Stage 1).
This provides a cost-effective recovery mechanism, saving significant LLM inference time compared to a full restart. 

\noindent{\bf Algorithm Mechanics.}
The algorithm maintains a current stage pointer (\textit{cur}) and a rollback target pointer (\textit{cycState}).
It executes stages sequentially from 1 to 4.
Upon a Stage 4 failure, the failure counter (\textit{regen}) increments, and the process rolls back to \textit{cycState} (initially Stage 3).
If failures persist (exceeding a threshold), \textit{cycState} is set to an earlier stage, enabling a deeper rollback.
This mechanism ensures that persistent errors in later stages can be addressed by regenerating earlier, potentially problematic components. 

\begin{algorithm}[h]
    \caption{Staged Rollback Algorithm for Harness Generation}
    \label{algo:generation}
    \begin{algorithmic}[1]
        \Require \(regen\), \(cur\), \(cycState\) and state union \(\{stage1, stage2, stage3, stage4\}\).
        \Ensure \(Harness\) code or FAIL.
        \State \(cur \leftarrow stage1\)
        \State \(cycState \leftarrow stage4\) \Comment{Initial rollback target}
        \State \(regen \leftarrow 0\)
        \While{\(True\)}
            \If{\(cur == stage1\)}
                \State \(genAPIDoc()\)
            \ElsIf{\(cur == stage2\)}
                \State \(genStructureCode()\)
            \ElsIf{\(cur == stage3\)}
                \State \(genRoughCode()\)
            \ElsIf{\(cur == stage4\)}
                \State \(code \leftarrow genHarnessCode()\)
                \State \(result \leftarrow compile\_and\_runCheck(code)\)
                \If{\(result\)}
                    \Return \(code\) \Comment{Success}
                \Else
                    \State \(regen \leftarrow regen + 1\)
                    \State \(cur \leftarrow cycState\) \Comment{Rollback to target stage}
                    \If{\(regen > 3\)} \Comment{Too many attempts}
                        \If{\(cycState == stage1\)}
                            \Return FAIL \Comment{Fail}
                        \EndIf
                        \State \(regen \leftarrow 0\)
                        \State \(cycState \leftarrow cycState - 1\) 
                        \Comment{Move rollback target earlier}
                    \EndIf
                \EndIf
            \EndIf
            \State \(cur \leftarrow cur + 1\) \Comment{Proceed to next stage}
        \EndWhile
    \end{algorithmic}
\end{algorithm}

\subsection{Implementation}
We implemented \tool in Python, using tree-sitter~\cite{Tree-sitter} for lightweight, robust syntax analysis. 
This approach trades off precise dataflow for generality, with subsequent LLM reasoning (Section~\ref{subsec:fa}) resolving ambiguities.
The codebase comprises approximately 4.7K lines of code and 139 prompt templates (modular components constituting our multi-stage pipeline; we list only the core templates in the Appendix~\ref{appendix4}, with the full set available in our  repository).
While our methodology is general, the current implementation targets C code.
Below, we highlight key implementation decisions that differentiate our approach.



\noindent{\bf Prompt Design Strategy.}
\tool uses LLMs for two distinct tasks: \textit{function identification} (in the grouping phase) and \textit{code generation} (in the harness phase).
For identification tasks (e.g., determining whether a parameter is a byte stream), validation is inherently difficult.
We therefore employ a three-prompt voting scheme: we craft three prompt variants (direct extraction, yes/no question, multiple-choice) for the same query and take the majority vote of the LLM's responses.
This significantly improves accuracy over single-prompt or repeated-same-prompt approaches.
For code generation tasks, we rely on syntactic validation and the rollback algorithm for error correction.

\noindent{\bf Validation of Intermediate Outputs.}
Each stage's output undergoes lightweight syntactic validation before proceeding.
We check for undefined functions or macros (e.g., calls to non-existent APIs), references to external libraries not in the project, and duplicate definitions.
The final harness from Stage 4 must additionally pass compilation and a 30‑second test (executing with empty input to ensure no immediate crash).
This layered validation catches errors early, prevents them from propagating to later stages, and is therefore essential for the rollback algorithm to function correctly.

\noindent{\bf Why Rollback, Not LLM-Based Repair?}
A deliberate design choice is to use staged rollback instead of prompting the LLM to ``fix'' erroneous code.
Our experiments show that LLM-based repair is often non-deterministic and can introduce new, unrelated errors while attempting to fix the original one.
The rollback mechanism provides a more reliable and efficient recovery path by reverting to a known-good earlier state and regenerating, leveraging the deterministic nature of our staged prompts.




\section{Evaluation}

In this section, we conducted comprehensive experiments to address the following research questions:

\begin{itemize}
    \item \textbf{RQ1 (Coverage):} How effective are \tool-generated harnesses in coverage?
    \item \textbf{RQ2 (Bug Detection Ability):} How effective are \tool-generated harnesses in triggering known bugs?
    \item \textbf{RQ3 (Real-world Bug Discovery):} Can \tool discover previously unreported bugs?
    \item \textbf{RQ4 (Efficiency):} How efficient is \tool compared to the competitors?
    \item \textbf{RQ5 (LLM Sensitivity):} How sensitive is \tool to the choice of LLM?
    \item \textbf{RQ6 (SFG Quality \& Ablation Study):} 
    What is the quality of SFG and how do SFG and staged decomposition mechanism individually contribute to overall effectiveness?
\end{itemize}

\noindent{\bf Environment.}

All experiments ran on an Ubuntu 24.04 node (Xeon Platinum 8160, 96 GB RAM, 4× NVIDIA V100 GPUs / 64 GB). 

\noindent{\bf Baselines.}
We selected OSS-Fuzz-Gen (git commit hash: 26b3259), CKGFuzzer (git commit hash: bb50d2f) and PromeFuzz (git commit 92df4c2) as baseline comparators. 
OSS-Fuzz-Gen, an LLM-based harness auto generation tool developed by Google, has been integrated into the OSS-Fuzz platform and is widely adopted in open-source projects. 
CKGFuzzer and PromeFuzz represents the current SOTA in academic research.
A functional comparison of these tools is summarized in Table~\ref{comp}.
To eliminate LLM choice as a confounding variable, all tools used the same backend model (Qwen3-32B) for the main experiments. 
For CKGFuzzer, feeding its entire call graph into the prompt caused token counts to exceed 2M, making generation infeasible (a limitation also noted in prior work~\cite{promefuzz}); we therefore truncated the call graph input to ensure a fair comparison.
We exclude traditional non‑LLM baselines because the transitive hierarchy (PromeFuzz > Hopper/libErator) is already known~\cite{promefuzz}.

\begin{table}[h]

\centering
\caption{A brief functional comparison with baselines.}
\label{comp}
\begin{tabular}{lcccc}
\toprule
\textbf{Approach} & \textbf{Scope} & \textbf{Target} & \textbf{\#Funcs} & \textbf{Feedback} \\
\hline
\rowcolor[HTML]{EFEFEF}
OSS-Fuzz-Gen & \textbf{All Funcs} & \textbf{Auto.} & Single & None \\
CKGFuzzer & API only & Manual & \textbf{Multiple} & None \\
\rowcolor[HTML]{EFEFEF}
PromeFuzz & API only & \textbf{Auto.} & \textbf{Multiple} & None \\
\tool & \textbf{All Funcs} & \textbf{Auto.} & \textbf{Multiple} & \textbf{Iterative} \\
\bottomrule
\end{tabular}
\end{table}

\begin{table*}[htbp]
  \centering
  \caption{
  Evaluation results of branch coverage (RQ1) and bug detection ability (RQ2). \#Br/\#Func: Total number of branches/functions, OF: OSS-Fuzz-Gen, CK: CKGFuzzer, PF: PromeFuzz, SF: \tool, \#Harness: Number of harnesses that trigger bugs.
   }
    \begin{tabular}{rrrrrrrc|cccc}
    \toprule
    \multicolumn{1}{c}{\multirow{2}[4]{*}{\textbf{Project}}}   & \multicolumn{1}{c|}{\multirow{2}[4]{*}{\textbf{Type}}} & \multicolumn{1}{c|}{\multirow{2}[4]{*}{\textbf{\#Br/\#Func}}} & \multicolumn{4}{c|}{\textbf{\#Covered Br/Func}} & \multicolumn{1}{c|}{\multirow{2}[4]{*}{\textbf{\makecell{Bug \\ ID}}}} & \multicolumn{4}{c}{\textbf{\#Harnesses}} \\
\cmidrule{4-7}\cmidrule{9-12}                  & \multicolumn{1}{c|}{} & \multicolumn{1}{c|}{} & \multicolumn{1}{c}{\textbf{OF}} & \multicolumn{1}{c}{\textbf{CK}} & \multicolumn{1}{c}{\textbf{PF}} & \multicolumn{1}{c|}{\textbf{SF}} &       & \textbf{OF} & \textbf{CK} & \textbf{PF} & \textbf{SF} \\
    \midrule
    \rowcolor[HTML]{EFEFEF}
    \multicolumn{1}{c}{c-ares}   & \multicolumn{1}{c|}{Lib} & \multicolumn{1}{c|}{8714 / 534} & \multicolumn{1}{c}{3914 / 69} & \multicolumn{1}{c}{3547 / 89} & \multicolumn{1}{c}{3848 / 112} & \multicolumn{1}{c|}{\textbf{4299} / 66} & \multicolumn{1}{l|}{CVE-2020-22217} & 5     & 4     & 7     & \textbf{9} \\
    \multicolumn{1}{c}{fribidi}   & \multicolumn{1}{c|}{Lib} & \multicolumn{1}{c|}{1748 / 60} & \multicolumn{1}{c}{727 / 30} & \multicolumn{1}{c}{465 / 15} & \multicolumn{1}{c}{774 / 16} & \multicolumn{1}{c|}{\textbf{1039} / 16} & \multicolumn{1}{l|}{CVE-2022-25310} & 4     & 5     & \textbf{8} & \textbf{8} \\
    \rowcolor[HTML]{EFEFEF}
    \multicolumn{1}{c}{libyaml}   & \multicolumn{1}{c|}{Lib} & \multicolumn{1}{c|}{6402 / 58} & \multicolumn{1}{c}{2671 / 23} & \multicolumn{1}{c}{3099 / 38} & \multicolumn{1}{c}{3078 / 57} & \multicolumn{1}{c|}{\textbf{4020} / 49} & \multicolumn{1}{l|}{issue:42486502} & 7     & 8     & 7     & \textbf{8} \\
    \multicolumn{1}{c}{cjson}   & \multicolumn{1}{c|}{Lib} & \multicolumn{1}{c|}{1546 / 92} & \multicolumn{1}{c}{699 / 25} & \multicolumn{1}{c}{670 / 27} & \multicolumn{1}{c}{792 / 78} & \multicolumn{1}{c|}{\textbf{1117} / 78} & \multicolumn{1}{l|}{CVE-2024-31755} & 7     & 8     & 7     & \textbf{10} \\
    \rowcolor[HTML]{EFEFEF}
    \multicolumn{1}{c}{sqlite3}   & \multicolumn{1}{c|}{Lib} & \multicolumn{1}{c|}{40236 / 2432} & \multicolumn{1}{c}{16879 / 39} & \multicolumn{1}{c}{19874 / 74} & \multicolumn{1}{c}{22580 / 246} & \multicolumn{1}{c|}{\textbf{28063} / 528} & \multicolumn{1}{l|}{CVE-2020-13434} & 5     & 7     & 8     & \textbf{9} \\
    \multicolumn{1}{c}{zlib}  & \multicolumn{1}{c|}{Lib} & \multicolumn{1}{c|}{2670 / 209} & \multicolumn{1}{c}{543 / 75} & \multicolumn{1}{c}{1436 / 69} & \multicolumn{1}{c}{1739 / 91} & \multicolumn{1}{c|}{\textbf{1810} / 68} & \multicolumn{1}{l|}{CVE-2013-0899} & 2     & 3     & 5     & \textbf{9} \\
    \rowcolor[HTML]{EFEFEF}
    \multicolumn{1}{c}{opus}  & \multicolumn{1}{c|}{Lib} & \multicolumn{1}{c|}{11470 / 532} & \multicolumn{1}{c}{1422 / 8} & \multicolumn{1}{c}{1789 / 12} & \multicolumn{1}{c}{4556 / 53} & \multicolumn{1}{c|}{\textbf{7296} / 143} & \multicolumn{1}{l|}{CVE-2024-47607} & 5     & 7     & 7     & \textbf{8} \\
    \multicolumn{1}{c}{libxml2}   & \multicolumn{1}{c|}{Lib} & \multicolumn{1}{c|}{59620 / 973} & \multicolumn{1}{c}{5831 / 160} & \multicolumn{1}{c}{4876 / 65} & \multicolumn{1}{c}{1965 / 88} & \multicolumn{1}{c|}{\textbf{6316} / 76} & \multicolumn{1}{l|}{CVE-2025-27113} & 4     & 5     & 6     & \textbf{9} \\
    \rowcolor[HTML]{EFEFEF}
    \multicolumn{1}{c}{lz4}   & \multicolumn{1}{c|}{Lib} & \multicolumn{1}{c|}{1964 / 188} & \multicolumn{1}{c}{489 / 59} & \multicolumn{1}{c}{552 / 48} & \multicolumn{1}{c}{874 / 60} & \multicolumn{1}{c|}{\textbf{1167} / 45} & \multicolumn{1}{l|}{CVE-2023-35955} & 3     & 6     & 7     & \textbf{10} \\
    \multicolumn{1}{c}{libssh2}  & \multicolumn{1}{c|}{Lib} & \multicolumn{1}{c|}{6184 / 347} & \multicolumn{1}{c}{88 / 79} & \multicolumn{1}{c}{96 / 22} & \multicolumn{1}{c}{66 / 26} & \multicolumn{1}{c|}{\textbf{172} / 25} & \multicolumn{1}{l|}{CVE-2020-22218} & 3     & 5     & 6     & \textbf{10} \\
    \rowcolor[HTML]{EFEFEF}
    \multicolumn{1}{c}{expat}   & \multicolumn{1}{c|}{Lib} & \multicolumn{1}{c|}{6650 / 244} & \multicolumn{1}{c}{2598 / 26} & \multicolumn{1}{c}{3348 / 52} & \multicolumn{1}{c}{3720 / 76} & \multicolumn{1}{c|}{\textbf{4246} / 48} & \multicolumn{1}{l|}{CVE-2022-40674} & 2     & 1     & 4     & \textbf{8} \\
    \multicolumn{1}{c}{avahi}   & \multicolumn{1}{c|}{Lib} & \multicolumn{1}{c|}{5506 / 815} & \multicolumn{1}{c}{654 / 42} & \multicolumn{1}{c}{668 / 44} & \multicolumn{1}{c}{288 / 99} & \multicolumn{1}{c|}{\textbf{993} / 41} & \multicolumn{1}{l|}{CVE-2023-38473} & 7     & 6     & \textbf{8} & \textbf{8} \\
    \rowcolor[HTML]{EFEFEF}
   \multicolumn{1}{c}{libxslt}  & \multicolumn{1}{c|}{Lib} & \multicolumn{1}{c|}{42964 / 155} & \multicolumn{1}{c}{5741 / 145} & \multicolumn{1}{c}{5411 / 86} & \multicolumn{1}{c}{5275 / 198} & \multicolumn{1}{c|}{\textbf{9190} / 88} & \multicolumn{1}{l|}{CVE-2024-55549} & 1     & 1     & 3     & \textbf{8} \\
    \multicolumn{1}{c}{libtiff}   & \multicolumn{1}{c|}{Lib} & \multicolumn{1}{c|}{14328 / 319} & \multicolumn{1}{c}{4698 / 102} & \multicolumn{1}{c}{4896 / 107} & \multicolumn{1}{c}{5290 / 193} & \multicolumn{1}{c|}{\textbf{5906} / 165} & \multicolumn{1}{l|}{CVE-2023-6277} & 4     & 3     & 5     & \textbf{7} \\
    \rowcolor[HTML]{EFEFEF}
    \multicolumn{1}{c}{lcms} & \multicolumn{1}{c|}{Lib} & \multicolumn{1}{c|}{8956 / 487} & \multicolumn{1}{c}{2578 / 165} & \multicolumn{1}{c}{2887 / 138} & \multicolumn{1}{c}{\textbf{3424} / 358} & \multicolumn{1}{c|}{3256 / 277} & \multicolumn{1}{l|}{CVE-2025-29069} & 6     & 7     & \textbf{8} & \textbf{8} \\
    \multicolumn{1}{c}{libarchive}& \multicolumn{1}{c|}{Lib} & \multicolumn{1}{c|}{21204 / 637} & \multicolumn{1}{c}{576 / 52} & \multicolumn{1}{c}{4687 / 98} & \multicolumn{1}{c}{6446 / 382} & \multicolumn{1}{c|}{\textbf{8313} / 146} & \multicolumn{1}{l|}{CVE-2025-5914} & 6     & 7     & 7     & \textbf{10} \\
    \rowcolor[HTML]{EFEFEF}
    \multicolumn{1}{c}{xz}   & \multicolumn{1}{c|}{Lib} & \multicolumn{1}{c|}{2672 / 323} & \multicolumn{1}{c}{968 / 42} & \multicolumn{1}{c}{876 / 48} & \multicolumn{1}{c}{1356 / 78} & \multicolumn{1}{c|}{\textbf{1955} / 120} & \multicolumn{1}{l|}{CVE-2025-31115} & 4     & 2     & 6     & \textbf{8} \\
    \multicolumn{1}{c}{jq}   & \multicolumn{1}{c|}{App} & \multicolumn{1}{c|}{18942 / 342} & \multicolumn{1}{c}{596 / 13} & \multicolumn{1}{c}{N/A} & \multicolumn{1}{c}{235 / 12} & \multicolumn{1}{c|}{\textbf{2909} / 28} & \multicolumn{1}{l|}{CVE-2025-48060} & 4     & 5     & 8     & \textbf{9} \\
    \rowcolor[HTML]{EFEFEF}
    \multicolumn{1}{c}{pjsip}  & \multicolumn{1}{c|}{App} & \multicolumn{1}{c|}{41674 / 2980} & \multicolumn{1}{c}{98 / 9} & \multicolumn{1}{c}{N/A} & \multicolumn{1}{c}{145 / 104} & \multicolumn{1}{c|}{\textbf{482} / 198} & \multicolumn{1}{l|}{CVE-2023-27585} & 6     & 5     & 6     & \textbf{7} \\
    \multicolumn{1}{c}{kamailio}  & \multicolumn{1}{c|}{App} & \multicolumn{1}{c|}{143066 / 2494} & \multicolumn{1}{c}{887 / 58} & \multicolumn{1}{c}{N/A} & \multicolumn{1}{c}{112 / 143} & \multicolumn{1}{c|}{\textbf{2024} / 230} & \multicolumn{1}{l|}{CVE-2025-12206} & 6     & 1     & 5     & \textbf{7} \\
    \rowcolor[HTML]{EFEFEF}
    \multicolumn{1}{c}{postfix}  & \multicolumn{1}{c|}{App} & \multicolumn{1}{c|}{5778 / 1526} & \multicolumn{1}{c}{631 / 53} & \multicolumn{1}{c}{N/A} & \multicolumn{1}{c}{52 / 69} & \multicolumn{1}{c|}{\textbf{1286} / 540} & \multicolumn{1}{l|}{issue:42488602} & 3     & 2     & 5     & \textbf{8} \\
    \multicolumn{1}{c}{gdbm}   & \multicolumn{1}{c|}{App} & \multicolumn{1}{c|}{2204 / 156} & \multicolumn{1}{c}{205 / 20} & \multicolumn{1}{c}{N/A} & \multicolumn{1}{c}{96 / 39} & \multicolumn{1}{c|}{\textbf{668} / 114} & \multicolumn{1}{l|}{issue:42533836} & 7     & 6     & 6     & \textbf{7} \\
    \rowcolor[HTML]{EFEFEF}
    \multicolumn{1}{c}{file}  & \multicolumn{1}{c|}{App} & \multicolumn{1}{c|}{7886 / 246} & \multicolumn{1}{c}{765 / 111} & \multicolumn{1}{c}{N/A} & \multicolumn{1}{c}{1196 / 114} & \multicolumn{1}{c|}{\textbf{2653} / 166} & \multicolumn{1}{l|}{issue:391975635} & 8     & 6     & 7     & \textbf{8} \\
    \multicolumn{1}{c}{hpn-ssh}  & \multicolumn{1}{c|}{App} & \multicolumn{1}{c|}{8832 / 1661} & \multicolumn{1}{c}{55 / 20} & \multicolumn{1}{c}{N/A} & \multicolumn{1}{c}{44 / 164} & \multicolumn{1}{c|}{\textbf{415} / 495} & \multicolumn{1}{l|}{issue:371061096} & 5     & 4     & 4     & \textbf{7} \\
    \rowcolor[HTML]{EFEFEF}
    \multicolumn{1}{c}{dropbear}  & \multicolumn{1}{c|}{App} & \multicolumn{1}{c|}{8012 / 432} & \multicolumn{1}{c}{34 / 22} & \multicolumn{1}{c}{N/A} & \multicolumn{1}{c}{56 / 62} & \multicolumn{1}{c|}{\textbf{118} / 27} & \multicolumn{1}{l|}{issue:391975635} & 8     & 7     & 8     & 9 \\
    \midrule
          &           &          &       &       &       &       & \textbf{Aver. Rate(\%)} & 46.9 & 55.0 & 61.3 & \textbf{83.3} \\
\cmidrule{8-12}     \end{tabular}%

  \label{expr1}%
\end{table*}%

\noindent{\bf Dataset.} We select 25 open-source C projects from GitHub (17 libraries, 8 applications; details in Appendix~\ref{app:rq1_detail}, Table~\ref{project_detail}). Our dataset design follows three principles to ensure rigorous and unbiased evaluation:
\begin{itemize}
    \item \textbf{Benchmarking:} We include 6 established targets (c-ares, cjson, zlib, libtiff, lcms, sqlite3) commonly used in prior work~\cite{promefuzz,xu2024ckgfuzzer,Hopper,liu2024afgen,lyu2024prompt}, ensuring direct comparability.
    \item \textbf{Stress Testing:} We incorporate projects with complex protocol or file-parsing logic to push the boundaries of current generation methods.
    \item \textbf{Mitigating Data Contamination:} We introduce 11 libraries never studied in related work. Since LLMs may have seen harnesses for popular libraries, evaluating on these novel targets better isolates the intrinsic performance of the generation techniques themselves.
\end{itemize}
Additionally, we include 8 applications to test tools beyond their original library-focused scope (CKGFuzzer and PromeFuzz were designed solely for libraries).

\noindent{\bf Experimental Setup.}
All generated harnesses were compiled with LLVM and executed with libFuzzer; branch coverage was measured using \texttt{llvm-cov show} (Appendix~\ref{Collect_branch_op}). 
All fuzzing campaigns started from empty seeds (no initial corpus). 
Each harness ran for 24 hours over 5 trials (averaged and then rounded to the nearest integer) on a dedicated CPU core with no additional scheduling—each harness was tested independently immediately after successful generation. 
Generation settings varied by research question: for RQ1 (coverage), RQ5 (LLM sensitivity), and RQ6 (SFG quality \& ablation study), we generated one harness per function triplet (FT); for RQ2 (bug detection), we generated ten distinct harnesses per target bug function (Section~\ref{subsec:rq2}); for RQ3 (real‑world bug discovery), we generated ten harnesses per FT; and for RQ4 (efficiency), we measured generation time and token consumption only, without performing fuzzing. 
Our experimental design follows the prudent evaluation practices recommended for fuzzing research~\cite{schloegel2024sok}: standardized metrics, uniform fuzzing budgets (24 hours), cross‑tool comparison on a unified backend, and full reporting of both coverage and bug discovery results.

\subsection{RQ1: Coverage}
\label{subsec:rq1m}

\subsubsection{Experimental Setup} 
We tested 25 source projects by generating harnesses with each tool and running every harness for 24 hours from empty seeds. For each project, we aggregated the coverage data from all harnesses and computed the overall coverage.

For branch coverage, we collected dynamic execution traces using \texttt{llvm-cov}. All tools utilize libFuzzer as the backend in their implementations. The specific operations for collecting branch coverage are detailed in Appendix~\ref{Collect_branch_op}.
This setup differs from PromeFuzz's approach, which relies on the AFL-LTO backend. Notably, while PromeFuzz claims to utilize \texttt{llvm-cov} in their paper~\cite{promefuzz}, our inspection of their source code reveals that they actually employ \texttt{llvm-cov} in gcov compatibility mode to collect branch coverage\footnote{\url{https://github.com/pvz122/PromeFuzz/blob/master/database/utils/gcov.py:219}}. 

For function coverage, we extracted all callable functions exposed in the header files as the total function set. For each generated harness, we identified the functions directly invoked within it as the covered set.

\subsubsection{Metrics}

We quantified the relative improvement of \tool over each baseline as the mean ratio of branch coverage and function coverage across all projects. 
For each project $i$ and baseline $B$, we computed $R_{i,B}=\text{BrCov}(\tool_i)/\text{BrCov}(B_i)$ and $R_{i,F}=\text{FuncCov}(\tool_i)/\text{FuncCov}(B_i)$, then took the average of $R_{i,B}$ and $R_{i,F}$ over all projects $i$.

\subsubsection{Results}
The coverage evaluation results are shown in Table~\ref{expr1} (the \textit{\#Covered Br/Func} column).
\tool outperforms OSS-Fuzz-Gen, CKGFuzzer, and PromeFuzz in 24 out of 25 projects in branch coverage. 
Overall, \tool achieves 3.07$\times$, 1.71$\times$, and 4.26$\times$ higher branch coverage than OSS-Fuzz-Gen, CKGFuzzer, and PromeFuzz, respectively. For function coverage, \tool achieves 4.97$\times$, 2.33$\times$, and 1.52$\times$ higher coverage than OSS-Fuzz-Gen, CKGFuzzer, and PromeFuzz, respectively.

Specifically, for libraries, \tool improves branch coverage by factors of 2.64$\times$, 1.71$\times$, and 1.60$\times$ over OSS-Fuzz-Gen, CKGFuzzer, and PromeFuzz.
For applications, where CKGFuzzer is inapplicable, \tool achieves 3.98$\times$ and 9.9$\times$ higher coverage compared to OSS-Fuzz-Gen and PromeFuzz.

Notably, although baselines occasionally cover a larger number of functions in some individual projects, their branch coverage remains substantially lower. 
Manual inspection of baseline-generated harnesses reveals that many perform superficial invocations without properly routing external fuzz inputs (\texttt{data} parameter) into the target functions. 
Consequently, they achieve high function counts but fail to exercise internal control-flow paths. 
In contrast, \tool's SFG-driven generation explicitly models data dependencies, ensuring that harnesses correctly channel external inputs into target functions. 
This data-aware construction enables deeper logical exploration and explains why \tool consistently triggers more branches despite occasionally covering fewer functions.

This indicates that \tool-generated harnesses exhibit superior path discovery capability.
Our manual analysis reveals three key factors contributing to this performance gap:

\noindent{\bf Comprehensive Function Selection.}
The superior coverage of \tool stems from identifying a more complete set of relevant functions through deep dataflow aggregation. 
While PromeFuzz relies on heuristics based on direct code consumption (lacking transitivity), \tool constructs an SFG to model global function relationships, capturing transitive dependencies and long-range data flows that baselines systematically miss. 
CKGFuzzer's API-only focus and OSS-Fuzz-Gen's single-function approach further restrict their achievable coverage.

\noindent{\bf Staged Generation Process.}
The quality and reliability of our generated harnesses are ensured by the staged generation pipeline.
This contrasts sharply with the monolithic generation paradigm of the baselines.
By breaking the complex task into incremental stages with validation at each step, we transform a high-risk process into a series of simple, verifiable steps.
The baselines' monolithic approach is fragile; a single incorrect API usage can invalidate the entire output, yielding low-quality code.

\noindent{\bf Input Function Isolation.}
\tool maintains a one-to-one correspondence between ISFs and harnesses, whereas CKGFuzzer and PromeFuzz often combine multiple ISFs into a single harness. 
When these functions process divergent input formats, the fuzzer's genetic algorithm may skew coverage and reduce effectiveness—as observed in c-ares where merging \texttt{ares\_parse\_soa\_reply} and \texttt{ares\_parse\_aaaa\_reply} diminished coverage feedback specificity.

The primary reason for \tool's slightly lower coverage on \texttt{lcms} compared to PromeFuzz stems from our use of a tree-sitter-based parser.
\texttt{lcms} uses complex macro definitions that can confuse syntax-based analysis, causing some functions to be missed during extraction.
Five protocol implementations (\texttt{hpn-ssh}, \texttt{pjsip}, \texttt{libssh2}, \texttt{dropbear}, \texttt{proftpd}) presented challenges for all tools due to strict input validation and cryptographic requirements.

This delineates the effective scope of our approach: \tool excels at testing data-processing libraries with clear structural dataflow, while stateful protocols with complex validation logic represent a known boundary for fuzz driver-based testing in general.

\begin{center}
\fcolorbox{black}{gray!10}{\parbox{.95\linewidth}{\textbf{Answer to RQ1}: \tool achieves superior branch coverage by systematically generating high-quality harnesses for a broader range of functions. This stems from two key innovations: a principled function classification and grouping mechanism based on deep dataflow analysis, and a robust staged decomposition and rollback process that ensures code reliability.}}
\end{center}

\subsection{RQ2: Bug Detection Ability}
\label{subsec:rq2}

\subsubsection{Experimental Setup} 
We selected 25 known bugs comprising 19 representative CVEs and 6 documented OSS-Fuzz issues~\cite{oss-fuzz-tracer} as our evaluation targets. 

For each tool, we specified the target bug functions, generated corresponding harnesses, then executed them to verify bug-triggering capability: for \tool, we identified the smallest FTs containing the target bug functions to minimize interference; for OSS-Fuzz-Gen, we directly provided the target bug functions; for CKGFuzzer, we selected from its output API list those interfaces that either contained or might invoke the bug functions; and for PromeFuzz, we identified function sets containing the target bug functions.

For each bug function, we generated 10 compilable harnesses per tool and executed them for 24 hours. 
Manual crash analysis then confirmed successful bug triggering.

\subsubsection{Metrics} We calculated the average trigger rate (Aver. Rate) as the fraction of harnesses successfully triggering a bug over the total allocated quota (10 $\times$ \#targets)
 for comparison with baselines.

\subsubsection{Results}
Table~\ref{expr1} (the \#Harness column) demonstrates that \tool's harnesses achieved significantly higher average bug detection rates.
Quantitatively, \tool achieves 1.77$\times$, 1.51$\times$, and 1.36$\times$ higher average trigger rates than OSS-Fuzz-Gen, CKGFuzzer, and PromeFuzz, respectively.
Our manual analysis reveals the following factors:

\noindent{\bf Robust Parameter Initialization.}
These baseline tools often produce harnesses with incorrect parameter initialization, hindering crash reproduction.
\tool addresses this through its staged approach, where Stage 1 handles independent parameter initialization and subsequent stages refine it (detailed in Section~\ref{subsec:stage-design}).

\noindent{\bf Comprehensive Verification.}
We observed that baseline harnesses often omitted critical function calls (LLM forgetting) or redundantly defined target functions (masking the real function).
\tool enforces tree-sitter-based syntactic and semantic validation at each generation stage, ensuring functional completeness.

\noindent{\bf Optimal Function Aggregation.}
PromeFuzz and CKGFuzzer produce harnesses containing multiple ISFs, which reduces bug detection efficiency.
\tool's one-ISF-per-harness strategy avoids this pitfall.

\begin{center}
\fcolorbox{black}{gray!10}{\parbox{.95\linewidth}{\textbf{Answer to RQ2}: \tool's staged generation with multi-stage validation ensures proper function invocation, leading to significantly higher bug detection rates than baselines.}}
\end{center}

\subsection{RQ3: Real-world Bug Discovery}
\label{subsec:rq3}

\subsubsection{Experimental Setup} We generated 10 validated harnesses per native target unit for each tool (FTs for \tool, function sets for PromeFuzz, API combinations for CKGFuzzer, single functions for OSS-Fuzz-Gen), and fuzzed each for 24 hours with an empty seed corpus.
Our crash analysis pipeline employed: 1) automated filtering to remove crashes in harness code, 2) call stack-based deduplication using libFuzzer traces to identify unique crash sites and 3) manual inspection to discard false positives (including API misuse cases), with true bugs confirmed through expert review. 
The manual verification revealed the following crash discovery results and previously unreported bugs.

\subsubsection{Results} 

Table~\ref{expr3-total} summarizes the crash discovery results. After crash call-stack deduplication, \tool obtained 40 unique crashes. Manual analysis classified 9 as false positives stemming from API misuse, and 31 as genuine bugs, resulting in a 77.5\% precision.
For comparison, PromeFuzz had a precision of 42.9\%, and CKGFuzzer and OSS-Fuzz-Gen achieved precisions of 5.6\% and 2.1\%, confirming \tool's superior reliability. This low false-positive rate is primarily due to our dataflow‑aware function grouping, which prevents irrelevant API combinations, and the staged rollback validation, which filters out harnesses that misuse APIs before fuzzing.

\begin{table}[htbp]
  \centering
  \caption{Crash discovery results from OSS-Fuzz-Gen, CKGFuzzer, PromeFuzz and \tool.}
    \begin{tabular}{cl|cccc}
    \toprule
     & \textbf{Type}  & \textbf{OF}    & \textbf{CK}    & \textbf{PF}    & \textbf{SF} \\
    \midrule
    \multirow{3}[1]{*}{FP} & API Contract Violations & 28    & 15    & 17    & 2 \\
          & Resource Lifecycle Mismatches & 12    & 14    & 8     & 4 \\
          & Unsafe Memory Operations  & 6     & 5     & 3     & 3 \\
    \midrule
    \multirow{3}[2]{*}{TP} & Bugs in Standard Functions & 0     & 0     & 0     & 7 \\
          & Bugs in Deprecated Functions  & 0     & 0     & 6     & 7 \\
          & Bugs in Unsafe Functions & 1     & 2     & 15    & 17 \\
    \bottomrule
          & \textbf{Precision Rate(\%)}    & 2.1  & 5.6 & 42.9 & \textbf{77.5} \\
    \cline{2-6}
    \end{tabular}%
  \label{expr3-total}%
\end{table}%

\begin{table}[htbp]
  \centering
  \caption{Bugs found by \tool-generated harness.}
    \begin{tabular}{c|ccc}
    \toprule
    \textbf{Project} & \textbf{Bug Type} & \multicolumn{1}{c}{\textbf{Bug ID}} & \textbf{Status} \\
    \midrule
    \rowcolor[HTML]{EFEFEF}
    expat &  Heap Overflow     &   -    & Reported \\
    lz4   &   Heap Overflow    &    -   & Reported \\
    \rowcolor[HTML]{EFEFEF}
    file & Stack Overflow & Commit id:acfbd73  & Fixed \\
    
    libarchive &   Infinite Loop    &  CVE-2025-61467   & Fixed \\
    \rowcolor[HTML]{EFEFEF}
    hpn-ssh & Stack Overflow      &   CVE-2025-51519   & Confirmed \\
    
    hpn-ssh &   Use After Free  &  CVE-2025-51521    & Confirmed \\
    \rowcolor[HTML]{EFEFEF}
    sqlite3 & Heap Overflow & CVE-2025-51644 & Confirmed \\
    \bottomrule
    \end{tabular}%
    \par\vspace{1em} 
  \label{expr3}%
\end{table}%

For \tool, among the 9 false positives, 2 violated documented parameter constraints (e.g., passing non-null-terminated byte streams to functions expecting C-strings, triggering crashes in \texttt{strlen}). 
4 resulted from incorrect resource management (e.g., double-free vulnerabilities caused by harnesses explicitly freeing resources already managed internally by the target API). 
The remaining 3 were harness-induced memory safety errors that manifested as crashes within the target code.

\tool identified 31 genuine bugs. Among them, 7 reside in deprecated (but still callable) functions, and 17 in functions explicitly marked as unsafe—i.e., requiring callers to guarantee parameter validity and discouraged for general use. We have reported all bugs in these two categories to the respective maintainers.

The remaining 7 bugs were submitted as standard bug reports; 5 have been confirmed, with 2 already patched, while none of the competing tools discovered any new bugs. %
These confirmed bugs are summarized in Table~\ref{expr3}.
Our analysis reveals that the vulnerable functions predominantly reside outside OSS-Fuzz's target set, explaining why they evaded detection despite approximately five years of continuous fuzzing~\cite{oss-fuzz-tracer}. 
\tool effectively generated valid harnesses for these overlooked APIs, uncovering bugs missed by existing infrastructure.

\subsubsection{Impact} 
All discovered bugs constitute high-severity risks. 
Maintainers responded promptly to our reports: patches for libarchive and file were merged shortly after disclosure, and 5 bugs have been assigned CVE IDs. 
These rapid remediations and official acknowledgments validate both the criticality of our findings and \tool's practical utility~\cite{schloegel2025confusing} in enabling responsible disclosure and timely patching of real-world vulnerabilities.

\subsubsection{Case Study} 
Listing \ref{expat_bug} shows a heap overflow bug in expat. 
In this case, \texttt{Function1} calls directly calls \texttt{VulnFunc} without checking the size of \texttt{mem}, \texttt{VulnFunc} writes \texttt{sizeof(struct normal\_encoding)} size to \texttt{mem} buffer without checking size of \texttt{mem} either. 
When the size of \texttt{mem} is less than \texttt{sizeof(struct \allowbreak normal\_encoding)} (line 10), heap overflow occurs.

\begin{lstlisting}[caption={A heap overflow bug in expat.}, label=expat_bug]
Function1(void *mem, int *table, CONVERTER convert,void *userData) {
  ENCODING *enc = VulnFunc(mem, table, convert, userData); 
  if (enc)
    ((struct normal_encoding *)enc)->type[ASCII_COLON] = BT_COLON;
  return enc;}
VulnFunc(void *mem,int *table,CONVERTER convert,void *userData){
  int i;
  struct unknown_encoding *e = (struct unknown_encoding *)mem;
  // crash here
  memcpy(mem,&latin1_encoding,sizeof(struct normal_encoding));
  ...}
\end{lstlisting}

Listing \ref{LZ4_bug} shows a heap overflow bug in lz4. In this case, the bug occurs during the handling of the final uncompressed literals (\texttt{last\_literals}) in the \texttt{VulnFunc} (line 9 and 12). A heap overflow is triggered when \texttt{LZ4\_memcpy} is executed due to incorrect calculations of the remaining space in the destination buffer. Because the \texttt{VulnFunc}'s parameter must be initialized correctly by \texttt{LZ4\_resetStream} function and only \tool can generate the harness that contains such call chain. Our case study demonstrates the effectiveness of \tool in detecting real-world bugs.

\begin{lstlisting}[caption={A heap overflow bug in lz4.}, label=LZ4_bug]
VulnFunc(LZ4_stream_t_internal* const cctx, const char* const source,...){
const BYTE* anchor = (const BYTE*) source;
...
_last_literals:
size_t lastRun = (size_t)(iend - anchor);
  if ( (outputDirective) && 
    (op + lastRun + 1 + ((lastRun+255-RUN_MASK)/255) > olimit)) {
      if (outputDirective == fillOutput) {
        lastRun  = (size_t)(olimit - op) - 1; // maybe negetive
        lastRun -= (lastRun + 256 - RUN_MASK) / 256;}}
...
LZ4_memcpy(op, anchor, lastRun); // crash here
\end{lstlisting}

\begin{center}
\fcolorbox{black}{gray!10}{\parbox{.95\linewidth}{\textbf{Answer to RQ3}: 
The results demonstrate that \tool's SFG-based function selection and staged rollback generation collectively enabled the discovery of 7 previously unreported bugs across 25 projects, conclusively validating its efficacy.}}
\end{center}

\subsection{RQ4: Efficiency}
\label{subsec:rq4m}

\subsubsection{Experimental Setup}
Based on the harness generation process in Section~\ref{subsec:rq1m}, we logged intermediate pipeline data across all 25 projects. We quantified the efficiency overhead of \tool by tracking generation latency and LLM resource consumption.

\subsubsection{Metrics}

The execution pipeline comprises two phases: project pre-processing and harness generation.
We report four metrics: PreT, the static pre-processing time per project in hours 
(reported as `-' for OSS-Fuzz-Gen because its target selection is interleaved with generation, not a separate preprocessing phase)
; GenT, the average generation time to produce one valid harness in hours; Tokens, the total LLM token consumption per project in millions, which directly correlates with GPU compute cycles under our local deployment; and \#S, the average number of pipeline stages executed per valid harness for \tool, quantifying the overhead of the staged-rollback mechanism.

\subsubsection{Results}
Table~\ref{expr4-eff} summarizes the mean efficiency metrics across all projects, with per-project distributions in Figure~\ref{expr-eff} and detailed efficiency statistics in Appendix~\ref{app:eff_details}.

\tool averages 3.84h for pre-processing, outperforming CKGFuzzer (4.33h) and PromeFuzz (6.64h) by replacing heavy static analysis pipelines (CodeQL, embedding extraction, LibClang) with lightweight tree-sitter parsing, reserving LLM invocation solely for ambiguous pointer resolution. During generation, \tool achieves a mean GenT of 0.33h and consumes 23.55M tokens, substantially lower than OSS-Fuzz-Gen (1.64h, 35.47M) and PromeFuzz (1.39h, 31.26M). Baselines rely on monolithic prompts that accumulate unbounded context (expanding up to 32K tokens), where excessive prompt length induces quadratic scaling in Time-To-First-Token of LLM latency, compounding token waste upon regeneration failures. \tool mitigates this by enforcing a strict 8K token cap per stage and localizing retries through staged rollback. This architecture avoids costly full restarts, maintaining controlled latency and token overhead despite a higher API call frequency. The average execution depth remains low at 7.78 stages (only 3.78 above the ideal 4), confirming that rollbacks are shallow and computationally inexpensive.

CKGFuzzer's lower token usage (17.54M) reflects its library-only scope rather than inherent efficiency. As Figure~\ref{expr-eff} illustrates, application targets inherently increase average generation time and token consumption on all tools. Isolating libraries for a controlled comparison, \tool averages 0.23h and 13.82M tokens, substantially outperforming OSS-Fuzz-Gen (0.75h, 19.65M), CKGFuzzer(0.94h, 17.54M) and PromeFuzz (0.95h, 14.92M). 
\tool achieves the lowest per-valid-harness cost across all scopes (see Appendix~\ref{app:eff_details} for detailed \#Harness numbers).
This confirms that the staged architecture delivers consistent efficiency gains while scaling robustly to complex applications, translating compute savings directly into higher-quality, coverage-driving outputs.

\begin{table}[t]
  \centering
  \caption{Average efficiency analysis results across all 25 target projects. 
 PreT: mean pre-processing time (hours); 
 GenT: mean generation time per harness (hours); 
 Tokens: mean LLM token consumption (millions); 
 \#S: mean number of generation stages per harness. 
 }
    \begin{tabular}{c|cccc}
    \toprule
    \textbf{Tool} & \textbf{PreT(h)} & \textbf{GenT(h)} & \textbf{Tokens(M)} & \textbf{\#S} \\
    \midrule
        \rowcolor[HTML]{EFEFEF}
    Oss-Fuzz-Gen & -     & 1.64  & 35.47 & -\\
    CKGFuzzer & 4.33  & 0.94   & 17.54 & - \\
        \rowcolor[HTML]{EFEFEF}
    PromeFuzz & 6.64  & 1.39  & 31.26 & -\\
    \tool & 3.84  & 0.33  & 23.55 & 7.78 \\
    \bottomrule
    \end{tabular}%
  \label{expr4-eff}%
\end{table}%

\begin{figure}[htbp]
    \centering
    \includegraphics[width=1.00\linewidth]{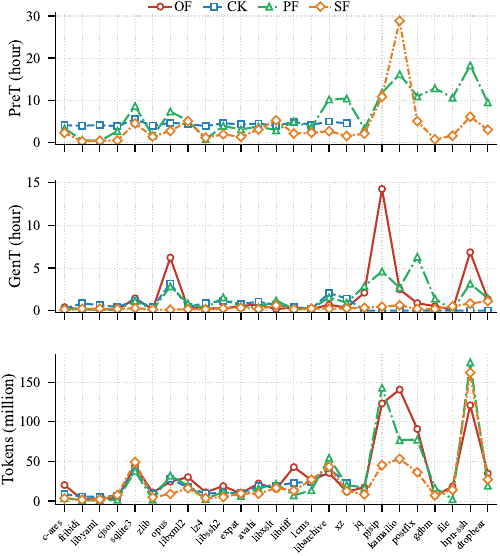}
    \caption{Efficiency metrics across 25 projects. PreT: pre-processing time (hours), GenT: average generation time per harness (hours), Tokens: LLM token consumption(millions). }
    \label{expr-eff}
\end{figure}

\begin{center}
\fcolorbox{black}{gray!10}{\parbox{.95\linewidth}{\textbf{Answer to RQ4}: \tool achieves competitive efficiency: lightweight tree-sitter parsing eliminates heavy static analysis overhead, while the staged rollback mechanism controls generation latency and token consumption by strictly capping per-stage context length and localizing retries. The modest stage count (7.78 vs. ideal 4) confirms that rollbacks remain shallow, introducing negligible cost while substantially improving harness validity.}}
\end{center}

\subsection{RQ5: LLM Sensitivity}
\label{subsec:rq4}

\subsubsection{Experimental Setup} Due to cost constraints, we selected four representative targets (comprising two libraries and two applications) to assess the impact of different LLM backends. Since \tool relies on a structured prompt template with only minor payload variations, the prompts from these targets provide sufficient coverage to reflect the backend’s overall influence. We evaluated performance using four models: Qwen3-32B (primary), Qwen3-14B (lower), DeepSeek-V3~\cite{deepseekv3} (intermediate), and Claude-3.5-Opus~\cite{claude3_5opus} (higher performance~\cite{gao2025comparison}), with each harness running continuously for 24 hours.

\subsubsection{Metrics} We analyzed the branch coverage of harnesses generated by different LLMs to evaluate how LLM selection affects the effectiveness of generated harnesses.

\begin{table}[htbp]
  \centering
  \caption{Branch coverage with different LLMs.}

\begin{tabular}{c|cccc}
\toprule
    \multicolumn{1}{c|}{\multirow{2}{*}{\textbf{Project}}} & \multicolumn{4}{c}{\textbf{\#Covered Branches}} \\
\cmidrule{2-5}
    & \textbf{Qwen14B} & \textbf{Qwen32B} & \textbf{DeepSeek} & \textbf{Claude} \\
\midrule

    \rowcolor[HTML]{EFEFEF}
    lz4   & 957 & 1167 & 1421 & \textbf{1445} \bigstrut[t]\\
    libyaml & 3211 & 3937 & 4231 & \textbf{4421} \\
    \rowcolor[HTML]{EFEFEF}
    proftpd & 58 & \textbf{120} & \textbf{120} & \textbf{120} \\
    dropbear & 110 & \textbf{118} & \textbf{118} & \textbf{118} \bigstrut[b]\\
\bottomrule
    \end{tabular}%
  \label{result-rq4}%
\end{table}%

\subsubsection{Results}
As shown in Table~\ref{result-rq4}, Qwen3-32B performs comparably to DeepSeek-V3 and Claude-3.5-Opus on two projects.
It slightly trails in two projects, primarily due to weaker enumeration type parameter selection.
However, branch coverage differences are marginal overall.
Qwen3-14B lags significantly in all metrics, attributed to its smaller size and inferior sub-task processing under the LLM scaling law.
The results suggest that once model scaling reaches a sufficient threshold, \tool's effectiveness becomes less dependent on raw model performance, as the decomposed subtasks are manageable.

\begin{center}
\fcolorbox{black}{gray!10}{\parbox{.95\linewidth}{\textbf{Answer to RQ5}: 
This result demonstrates that \tool's effectiveness becomes less dependent on model performance once a sufficient scale is reached, confirming that our decomposition processes effectively mitigate reliance on raw model capability.}}
\end{center}

\subsection{RQ6: SFG Quality \& Ablation Study}
\label{subsec:rq6}

\subsubsection{Experimental Setup}
We evaluate the structural quality of extracted SFGs and the individual contributions of SFG construction and the Staged Decomposition and Rollback (SDR) mechanism. For ablation, we construct two variants: SF-PF, which replaces our SFG with PromeFuzz's heuristic-based function grouping, and -SDR, which collapses the four-stage pipeline prompts into a single monolithic prompt (capped at 10 retries per target). All variants are evaluated under identical same fuzzing campaigns as in Section~\ref{subsec:rq1m}.

\subsubsection{Metrics}
SFG quality is measured by precision and the absolute number of omitted relevant functions:
\(\text{Precision} = \frac{\#\text{FT} - \#\text{FP}}{\#\text{FT}}\), where \(\#\text{FT}\) is the total number of extracted Function Triplets and \(\#\text{FP}\) is the number of false-positive FTs (FTs containing irrelevant functions). \(\#\text{O}\) denotes the number of relevant functions missing from the extracted SFG (function-level false negatives) — not the number of omitted FTs. A very small \(\#\text{O}\) implies near-complete coverage. Ablation performance is quantified as the relative branch coverage drop compared to the full \tool configuration.

\subsubsection{Results}

Table~\ref{expr5} summarizes the evaluation. Across all 25 projects, \tool extracts 11,326 FTs, of which 1,098 are false positives, giving a precision of 90.3\%. False positives primarily stem from the inherent limitations of static dataflow tracking, which captures parameter propagation but cannot verify semantic processing constraints. These FPs are naturally filtered during harness compilation and validation, thus not impacting fuzzing effectiveness. The overall false negative count is minimal, with only 19 relevant functions missed across all projects. These omissions are concentrated exclusively in \texttt{lcms} (7) and \texttt{dropbear} (12), where macro-dependent function names bypass our syntax-based parser.

Replacing our SFG with PromeFuzz function grouping (SF-PF) reduces branch coverage by 29.6\% relative to the full \tool configuration. Manual analysis reveals that heuristic grouping frequently omits critical ISFs, introduces irrelevant functions, and breaks transitive dataflow dependencies. Consequently, generated harnesses fail to correctly route external inputs through deep call chains. This confirms that \tool's SFG accurately models inter-procedural data dependencies, enabling harnesses to exercise complex internal logic. Notably, SF-PF still outperforms the original PromeFuzz, isolating the remaining performance gain to our SDR generation mechanism.

Disabling SDR (-SDR) causes a coverage decline of 46.7\% relative to the full \tool configuration. Generating a harness for a complete FT in a single monolithic prompt exceeds the LLM's effective reasoning capacity, leading to syntactic errors, incorrect API sequencing, and frequent generation failures. The staged pipeline decomposes this complex task into verifiable sub-steps, while localized rollback prevents cascading failures. This demonstrates that SDR is not merely an optimization but a structural necessity for synthesizing valid harnesses targeting deep dataflow sequences.

\begin{table}[htbp]
    \centering
    \caption{SFG quality and ablation results. 
    \#FP/\#FT: ratio of FTs containing irrelevant functions to total FTs.
    \#O: the number of relevant functions missing from the extracted SFG.
    SF-PF: variant replacing SFG with PromeFuzz's heuristic grouping; -SDR: variant without staged decomposition and rollback. Coverage drops ($\downarrow$) are relative to the full \tool configuration.}

    \begin{tabular}{c|cc|ccc}
        \toprule
        \multicolumn{1}{c|}{\multirow{2}{*}{\textbf{Project}}} & \multicolumn{2}{c|}{\textbf{SFG Quality}} & \multicolumn{3}{c}{\textbf{\#Covered Branches}} \\
        \cmidrule{2-6}
        & \textbf{\#FP/\#FT} & \textbf{\#O} & \textbf{SF} & \textbf{SF-PF} & \textbf{-SDR} \\
        \midrule
        \rowcolor[HTML]{EFEFEF}
        c-ares    & 7/323   & 0 & \textbf{4299} & 3787$ \downarrow${\scriptsize 12\%} & 2544$ \downarrow${\scriptsize 41\%} \\
        fribidi   & 2/30    & 0 & \textbf{999}  & 820$\downarrow${\scriptsize 18\%}  & 545$\downarrow${\scriptsize 45\%} \\
        \rowcolor[HTML]{EFEFEF}
        libyaml   & 0/36    & 0 & \textbf{4020} & 3200$\downarrow${\scriptsize 20\%} & 2152$\downarrow${\scriptsize 47\%} \\
        cjson     & 0/154   & 0 & \textbf{1117} & 830$\downarrow${\scriptsize 26\%}  & 449$\downarrow${\scriptsize 60\%} \\
        \rowcolor[HTML]{EFEFEF}
        sqlite3   & 18/626  & 0 & \textbf{28063}& 23765$\downarrow${\scriptsize 15\%}& 17373$\downarrow${\scriptsize 38\%} \\
        zlib      & 0/165   & 0 & \textbf{1810} & 1776$\downarrow${\scriptsize 2\%}   & 1115$\downarrow${\scriptsize 38\%} \\
        \rowcolor[HTML]{EFEFEF}
        opus      & 8/395   & 0 & \textbf{7296} & 5267$\downarrow${\scriptsize 28\%} & 4086$\downarrow${\scriptsize 44\%} \\
        libxml2   & 45/431 & 0 & \textbf{6316} & 2087$\downarrow${\scriptsize 67\%} & 3416$\downarrow${\scriptsize 46\%} \\
        \rowcolor[HTML]{EFEFEF}
        lz4       & 4/174   & 0 & \textbf{1167} & 730$\downarrow${\scriptsize 37\%}  & 494$\downarrow${\scriptsize 58\%} \\
        libssh2   & 13/145  & 0 & \textbf{172}  & 66$\downarrow${\scriptsize 62\%}   & 104$\downarrow${\scriptsize 40\%} \\
        \rowcolor[HTML]{EFEFEF}
        expat     & 0/173   & 0 & \textbf{4246} & 3987$\downarrow${\scriptsize 6\%}   & 1914$\downarrow${\scriptsize 55\%} \\
        avahi     & 16/278  & 0 & \textbf{993}  & 345$\downarrow${\scriptsize 65\%}  & 535$\downarrow${\scriptsize 46\%} \\
        \rowcolor[HTML]{EFEFEF}
        libxslt   & 0/130   & 0 & \textbf{9190} & 6763$\downarrow${\scriptsize 26\%} & 3685$\downarrow${\scriptsize 60\%} \\
        libtiff   & 13/452  & 0 & \textbf{5906} & 5376$\downarrow${\scriptsize 9\%}   & 2510$\downarrow${\scriptsize 58\%} \\
        \rowcolor[HTML]{EFEFEF}
        lcms      & 47/309  & 7 & \textbf{3256} & 3198$\downarrow${\scriptsize 2\%}   & 1614$\downarrow${\scriptsize 50\%} \\
        libarchive& 3/774   & 0 & \textbf{8313} & 5678$\downarrow${\scriptsize 32\%} & 4025$\downarrow${\scriptsize 52\%} \\
        \rowcolor[HTML]{EFEFEF}
        xz        & 1/188   & 0 & \textbf{1955} & 1543$\downarrow${\scriptsize 21\%} & 1228$\downarrow${\scriptsize 37\%} \\
        jq        & 38/195   & 0 & \textbf{2909} & 220$\downarrow${\scriptsize 92\%}  & 1359$\downarrow${\scriptsize 53\%} \\
        \rowcolor[HTML]{EFEFEF}
        pjsip     & 207/1748& 0 & \textbf{482}  & 109$\downarrow${\scriptsize 77\%}   & 303$\downarrow${\scriptsize 37\%} \\
        kamailio  & 392/1784& 0 & \textbf{2024} & 147$\downarrow${\scriptsize 93\%}  & 1244$\downarrow${\scriptsize 39\%} \\
        \rowcolor[HTML]{EFEFEF}
        postfix   & 223/1384& 0 & \textbf{1286} & 130$\downarrow${\scriptsize 90\%}  & 624$\downarrow${\scriptsize 52\%} \\
        gdbm      & 0/113   & 0 & \textbf{668}  & 149$\downarrow${\scriptsize 78\%}  & 316$\downarrow${\scriptsize 53\%} \\
        \rowcolor[HTML]{EFEFEF}
        file      & 0/226   & 0 & \textbf{2653} & 67$\downarrow${\scriptsize 97\%}   & 1092$\downarrow${\scriptsize 59\%} \\
        hpn-ssh   & 28/852 & 0 & \textbf{415}  & 33$\downarrow${\scriptsize 92\%}   & 261$\downarrow${\scriptsize 37\%} \\
        \rowcolor[HTML]{EFEFEF}
        dropbear  & 33/241  & 12& \textbf{118}  & 76$\downarrow${\scriptsize 36\%}   & 62$\downarrow${\scriptsize 48\%} \\
        \bottomrule
    \end{tabular}
    \label{expr5}
\end{table}

\begin{center}
\fcolorbox{black}{gray!10}{\parbox{.95\linewidth}{\textbf{Answer to RQ6}: 
SFG extraction achieves high fidelity (90.3\% precision filtered at compilation; only 19 global function-level omissions).
Ablation proves both components indispensable: the SFG captures transitive dataflows missed by heuristics, while SDR prevents LLM reasoning collapse on complex sequences, jointly enabling robust, coverage-driving harness synthesis.}}
\end{center}

\section{Discussion}

\subsection{Function Classification Sufficiency}
As defined in Section~\ref{sec:problem-definition}, our three categories (ISF, PRF, HPF) target functions involved in input-driven dataflow. They suffice to characterize all functions that process fuzzer inputs within \tool's intended scope (C projects with identifiable ISFs and typed struct propagation). Pure output functions (e.g., \texttt{printf}) are excluded as irrelevant. This classification has proven effective, enabling \tool to generate high-quality harnesses for a wide range of data-processing libraries and applications.

\subsection{Comparison with Non-LLM Approaches}
\tool inherits the generality of LLM-based generation while mitigating its key drawbacks (hallucinations, error cascades) via dataflow-aware aggregation and staged rollback. Compared to traditional non-LLM methods (e.g., consumer-code slicing~\cite{FUDGE,2020FuzzGen}, binary analysis~\cite{2021WINNIE,zhang2021apicraft}, dynamic API graphs~\cite{lin2025automatic}), \tool offers distinct advantages. Traditional methods are deterministic and resource-efficient, but their rigid synthesis often fails to capture long-range dataflow dependencies or generate semantically correct harnesses for complex libraries, resulting in low coverage on similar benchmarks. In contrast, \tool understands API semantics, infers usage patterns, and generates complex initialization sequences without relying on existing consumer code or unit tests. Consequently, \tool achieves significantly higher coverage and bug detection than both LLM-based and traditional baselines, offering a practical trade-off: an acceptably higher up-front computational cost for substantially improved effectiveness.

\subsection{Limitations}

\tool's current design entails several inherent limitations:

\noindent{\bf C++ Paradigm Support.} The framework focuses on C and lacks support for C++'s object-oriented constructs. Projects that initialize data via direct member assignment, rather than identifiable functions, also remain out of scope.

\noindent{\bf Semantic Validation.} Targets requiring strict, stateful semantic validation—such as protocol implementations and compiler components—present challenges, as their correctness conditions often exceed libFuzzer's input-space exploration model.

\noindent{\bf Limitations for Stateful Targets.}
Targets that require strict, multi-stage stateful interaction (e.g., protocol implementations) or complex input validation (e.g., compiler components) exceed the capabilities of libFuzzer's stateless, input-space exploration model.

Consequently, \tool is not designed for such targets and achieves low coverage on them like all other libFuzzer-based tools.

\section{Related Work}

\noindent{\textbf{Function-oriented Fuzzing.}}
Function-oriented fuzzing tests specific interfaces (e.g., libFuzzer~\cite{url_libfuzzer}) rather than entire programs, improving coverage through a divide-and-conquer strategy~\cite{crump2023libafl,guler2020cupid}.
Techniques automate this for diverse targets, from libraries (e.g., Hopper~\cite{Hopper}) to virtual machines (LibAFL QEMU~\cite{malmain2024libafl}), enhancing bug discovery~\cite{serebryany2016continuous}. 
Recent work continues to optimize, such as libErator~\cite{toffalini2025liberating} which balances resource allocation between driver generation and deep testing. 
A persistent challenge is understanding the project to select appropriate target functions. 
Our work addresses this via project digestion to annotate and group functions, improving target selection.

\noindent{\textbf{Automatic Harness Generation.}}
Given the pivotal role of harnesses, automated generation is crucial to reduce manual effort~\cite{FUDGE,2020FuzzGen,liu2024afgen,tran2021futag}. A common strategy analyzes runtime traces or consumer code to craft API sequences~\cite{2021WINNIE,zhang2021apicraft,jeong2023utopia}. 
For instance, Fudge~\cite{FUDGE} and FuzzGen~\cite{2020FuzzGen} synthesize harnesses from consumer code, while others like WINNIE~\cite{2021WINNIE} extend this to binaries. 
Recent approaches improve accuracy and diversity: NEXZZER~\cite{lin2025automatic} uses a dynamic API relation graph, WildSync~\cite{wu2025wildsync} recovers patterns from external projects, and POIROT~\cite{di2025hercules} targets Android native libraries. 
These methods require deep API usage understanding to construct valid sequences. 
Our approach employs LLMs for more accurate and automated generation without such reliance on external artifacts.


\noindent{\textbf{LLM-assisted Fuzzing.}}
LLMs are increasingly integrated into fuzzing for diverse tasks, from seed mutation to harness generation~\cite{zhang2024effective, deng2023large}.
For mutation-based fuzzing, Fuzz4All~\cite{xia2024fuzz4all} and WhiteFox~\cite{yang2023whitefox} employ GPT-4~\cite{achiam2023gpt} and StarCoder~\cite{li2023starcoder} as universal testers. 
For input generation, InputBlaster~\cite{liu2024testing} produces high-quality seeds via LLMs, ELFUZZ~\cite{ELFUZZ} synthesizes fuzzers through LLM-driven evolutionary search, and recent work~\cite{llmxps} generate complex inputs (e.g., XPS documents) through leverages LLM-assisted repair.
In library fuzzing, studies confirm LLM feasibility and coverage gains over traditional methods~\cite{zhang2023understanding, deng2023large}.
Dataflow-guided approaches like FlowFusion~\cite{llmphp} further show the value of leveraging data dependencies.
Specifically for harness generation, methods vary: PromptFuzz~\cite{lyu2024prompt} and OSS-Fuzz-Gen~\cite{Liu_OSS-Fuzz-Gen_Automated_Fuzz_2024} rely on function signatures; CKGFuzzer~\cite{xu2024ckgfuzzer} incorporates call graphs; PromeFuzz~\cite{promefuzz} builds a structured knowledge base for retrieval-augmented generation; and semantics-aware works include AIMFuzz~\cite{AIMFuzz} for binary analysis and OGHarn~\cite{sherman2025no} for oracle-guided synthesis.
Our work builds upon these advances but introduces a fundamental structural departure: a synthesis workflow whose stages are dictated by dataflow aggregation and which actively self-corrects via targeted rollback—a mechanism designed explicitly to contain the cascading errors typical of LLM-based code generation.
\section{Conclusion and Future Work}

In this paper, we presented \tool, a novel framework for automatically generating high-quality fuzz harnesses for C projects. It addresses the limitations of prior LLM-based methods through dataflow-aware function aggregation and a staged rollback-enabled workflow decomposition. Extensive evaluation on 25 real-world C projects shows that \tool substantially outperforms SOTA tools (OSS-Fuzz-Gen, CKGFuzzer, PromeFuzz)
, while maintaining acceptable efficiency. 
Crucially, \tool discovered 7 previously unreported bugs, demonstrating its practical effectiveness. Our work shows that integrating deep program analysis with a structured, fault-tolerant generation process unlocks LLMs' potential for complex software engineering tasks. Future work includes extending to 1) C++ via class parsing, 2) handling projects without clear ISFs, and 3) collecting workflow data for specialized fine-tuning.




\section*{Acknowledgment}

In this paper, we utilized DeepSeek solely for grammar refinement and stylistic polishing of pre-drafted content.



%

\bibliographystyle{unsrt}
\bibliography{software}

\begin{thebibliography}{10}

\bibitem{sokEternalWarinMemory}
Laszlo Szekeres, Mathias Payer, Tao Wei, and Dawn Song.
\newblock Sok: Eternal war in memory.
\newblock In {\em 2013 IEEE Symposium on Security and Privacy}, SP '13, page 48–62, USA, 2013. IEEE Computer Society.

\bibitem{manes2019art}
Valentin~JM Man{\`e}s, HyungSeok Han, Choongwoo Han, Sang~Kil Cha, Manuel Egele, Edward~J Schwartz, and Maverick Woo.
\newblock The art, science, and engineering of fuzzing: A survey.
\newblock {\em IEEE Transactions on Software Engineering}, 47(11):2312--2331, 2019.

\bibitem{serebryany2016continuous}
Kosta Serebryany.
\newblock Continuous fuzzing with libfuzzer and addresssanitizer.
\newblock In {\em 2016 IEEE Cybersecurity Development (SecDev)}, pages 157--157. IEEE, 2016.

\bibitem{fakhoury2024llm}
Sarah Fakhoury, Aaditya Naik, Georgios Sakkas, Saikat Chakraborty, and Shuvendu~K Lahiri.
\newblock Llm-based test-driven interactive code generation: User study and empirical evaluation.
\newblock {\em IEEE Transactions on Software Engineering}, 2024.

\bibitem{Liu_OSS-Fuzz-Gen_Automated_Fuzz_2024}
Dongge Liu, Oliver Chang, Jonathan metzman, Martin Sablotny, and Mihai Maruseac.
\newblock {OSS-Fuzz-Gen: Automated Fuzz Target Generation}, May 2024.

\bibitem{zhang2023understanding}
Cen Zhang, Mingqiang Bai, Yaowen Zheng, Yeting Li, Wei Ma, Xiaofei Xie, Yuekang Li, Limin Sun, and Yang Liu.
\newblock Understanding large language model based fuzz driver generation.
\newblock {\em arXiv e-prints}, pages arXiv--2307, 2023.

\bibitem{zhang2024effective}
Cen Zhang, Yaowen Zheng, Mingqiang Bai, Yeting Li, Wei Ma, Xiaofei Xie, Yuekang Li, Limin Sun, and Yang Liu.
\newblock How effective are they? exploring large language model based fuzz driver generation.
\newblock In {\em 33rd ACM SIGSOFT International Symposium on Software Testing and Analysis}, pages 1223--1235, 2024.

\bibitem{lyu2024prompt}
Yunlong Lyu, Yuxuan Xie, Peng Chen, and Hao Chen.
\newblock Prompt fuzzing for fuzz driver generation.
\newblock In {\em 2024 ACM SIGSAC Conference on Computer and Communications Security}, pages 3793--3807, 2024.

\bibitem{xu2024ckgfuzzer}
Hanxiang Xu, Wei Ma, Ting Zhou, Yanjie Zhao, Kai Chen, Qiang Hu, Yang Liu, and Haoyu Wang.
\newblock Ckgfuzzer: Llm-based fuzz driver generation enhanced by code knowledge graph.
\newblock In {\em 47th International Conference on Software Engineering: Companion Proceedings}, ICSE '25, page 243–254. IEEE Press, 2025.

\bibitem{gorz2025empirical}
Philipp G{\"o}rz, Joschua Schilling, Thorsten Holz, and Marcel B{\"o}hme.
\newblock An empirical study of fuzz harness degradation.
\newblock {\em arXiv preprint arXiv:2505.06177}, 2025.

\bibitem{huang2024understanding}
Xu~Huang, Weiwen Liu, Xiaolong Chen, Xingmei Wang, Hao Wang, Defu Lian, Yasheng Wang, Ruiming Tang, and Enhong Chen.
\newblock Understanding the planning of llm agents: A survey.
\newblock {\em arXiv preprint arXiv:2402.02716}, 2024.

\bibitem{promefuzz}
Yuwei Liu, Junquan Deng, Xiangkun Jia, Yanhao Wang, Minghua Wang, Lin Huang, Tao Wei, and Purui Su.
\newblock Promefuzz: A knowledge-driven approach to fuzzing harness generation with large language models.
\newblock In {\em 2025 ACM SIGSAC Conference on Computer and Communications Security}, CCS '25, page 1559–1573, New York, NY, USA, 2025. Association for Computing Machinery.

\bibitem{2021IntelliGen}
Mingrui Zhang, Jianzhong Liu, Fuchen Ma, Huafeng Zhang, and Yu~Jiang.
\newblock Intelligen: Automatic driver synthesis for fuzz testing.
\newblock In {\em 2021 IEEE/ACM 43rd International Conference on Software Engineering: Software Engineering in Practice (ICSE-SEIP)}, pages 318--327. IEEE, 2021.

\bibitem{FUDGE}
Domagoj Babić, Stefan Bucur, Yaohui Chen, Franjo Ivancic, Tim King, Markus Kusano, Caroline Lemieux, László Szekeres, and Wei Wang.
\newblock Fudge: fuzz driver generation at scale.
\newblock pages 975--985, 08 2019.

\bibitem{2020FuzzGen}
Kyriakos Ispoglou, Daniel Austin, Vishwath Mohan, and Mathias Payer.
\newblock $\{$FuzzGen$\}$: Automatic fuzzer generation.
\newblock In {\em 29th USENIX Security Symposium (USENIX Security 20)}, pages 2271--2287, 2020.

\bibitem{url_j40}
{j40}.
\newblock \url{https://github.com/lifthrasiir/j40 }, Accessed: 2024.

\bibitem{Tree-sitter}
T.~sitter Contributors.
\newblock {Tree-sitter: An incremental parsing system for programming tools}, 2025.

\bibitem{Hopper}
Peng Chen, Yuxuan Xie, Yunlong Lyu, Yuxiao Wang, and Hao Chen.
\newblock Hopper: Interpretative fuzzing for libraries.
\newblock pages 1600--1614, 11 2023.

\bibitem{liu2024afgen}
Yuwei Liu, Yanhao Wang, Xiangkun Jia, Zheng Zhang, and Purui Su.
\newblock Afgen: Whole-function fuzzing for applications and libraries.
\newblock In {\em 2024 IEEE Symposium on Security and Privacy (SP)}, pages 1901--1919. IEEE, 2024.

\bibitem{schloegel2024sok}
Moritz Schloegel, Nils Bars, Nico Schiller, Lukas Bernhard, Tobias Scharnowski, Addison Crump, Arash Ale-Ebrahim, Nicolai Bissantz, Marius Muench, and Thorsten Holz.
\newblock Sok: Prudent evaluation practices for fuzzing.
\newblock In {\em 2024 IEEE Symposium on Security and Privacy (SP)}, pages 1974--1993. IEEE, 2024.

\bibitem{oss-fuzz-tracer}
{OSS-Fuzz Tracker}.
\newblock \url{https://issues.oss-fuzz.com }, Accessed: Jan 2026.

\bibitem{schloegel2025confusing}
Moritz Schloegel, Daniel Klischies, Simon Koch, David Klein, Lukas Gerlach, Malte Wessels, Leon Trampert, Martin Johns, Mathy Vanhoef, Michael Schwarz, et~al.
\newblock Confusing value with enumeration: Studying the use of $\{$CVEs$\}$ in academia.
\newblock In {\em 34th USENIX Security Symposium (USENIX Security 25)}, pages 2887--2906, 2025.

\bibitem{deepseekv3}
DeepSeek-AI.
\newblock Deepseek-v3 technical report.
\newblock Technical report, DeepSeek-AI, 2025.

\bibitem{claude3_5opus}
Anthropic.
\newblock Claude 3.5 series update, 2024.

\bibitem{gao2025comparison}
Tianchen Gao, Jiashun Jin, Zheng~Tracy Ke, and Gabriel Moryoussef.
\newblock A comparison of deepseek and other llms.
\newblock {\em arXiv preprint}, arXiv:2502.03688, 2025.

\bibitem{2021WINNIE}
Jinho Jung, Stephen Tong, Hong Hu, Jungwon Lim, Yonghwi Jin, and Taesoo Kim.
\newblock Winnie: Fuzzing windows applications with harness synthesis and fast cloning.
\newblock In {\em 2021 Network and Distributed System Security Symposium (NDSS 2021)}, 2021.

\bibitem{zhang2021apicraft}
Cen Zhang, Xingwei Lin, Yuekang Li, Yinxing Xue, Jundong Xie, Hongxu Chen, Xinlei Ying, Jiashui Wang, and Yang Liu.
\newblock {{APICraft}}: Fuzz driver generation for closed-source {{SDK}} libraries.
\newblock In {\em 30th USENIX Security Symposium (USENIX Security 21)}, pages 2811--2828, 2021.

\bibitem{lin2025automatic}
Jiayi Lin, Qingyu Zhang, Junzhe Li, Chenxin Sun, Hao Zhou, Changhua Luo, and Chenxiong Qian.
\newblock Automatic library fuzzing through api relation evolvement.
\newblock In {\em 2025 Network and Distributed System Security Symposium (NDSS 2025)}, 2025.

\bibitem{url_libfuzzer}
{libfuzzer}.
\newblock \url{https://llvm.org/docs/LibFuzzer.html }, Accessed: Jan 2026.

\bibitem{crump2023libafl}
Addison Crump, Andrea Fioraldi, Dominik Maier, and Dongjia Zhang.
\newblock Libafl libfuzzer: Libfuzzer on top of libafl.
\newblock In {\em 2023 IEEE/ACM International Workshop on Search-Based and Fuzz Testing (SBFT)}, pages 70--72. IEEE, 2023.

\bibitem{guler2020cupid}
Emre G{\"u}ler, Philipp G{\"o}rz, Elia Geretto, Andrea Jemmett, Sebastian {\"O}sterlund, Herbert Bos, Cristiano Giuffrida, and Thorsten Holz.
\newblock Cupid: Automatic fuzzer selection for collaborative fuzzing.
\newblock In {\em 36th Annual Computer Security Applications Conference}, pages 360--372, 2020.

\bibitem{malmain2024libafl}
Romain Malmain, Andrea Fioraldi, and Aur{\'e}lien Francillon.
\newblock Libafl qemu: A library for fuzzing-oriented emulation.
\newblock In {\em BAR 2024, Workshop on Binary Analysis Research, colocated with NDSS 2024}, 2024.

\bibitem{toffalini2025liberating}
Flavio Toffalini, Nicolas Badoux, Zurab Tsinadze, and Mathias Payer.
\newblock Liberating libraries through automated fuzz driver generation: Striking a balance without consumer code.
\newblock volume~2, New York, NY, USA, June 2025. Association for Computing Machinery.

\bibitem{tran2021futag}
Chi~Thien Tran and Shamil Kurmangaleev.
\newblock Futag: Automated fuzz target generator for testing software libraries.
\newblock In {\em 2021 Ivannikov Memorial Workshop (IVMEM)}, pages 80--85. IEEE, 2021.

\bibitem{jeong2023utopia}
Bokdeuk Jeong, Joonun Jang, Hayoon Yi, Jiin Moon, Junsik Kim, Intae Jeon, Taesoo Kim, WooChul Shim, and Yong~Ho Hwang.
\newblock Utopia: Automatic generation of fuzz driver using unit tests.
\newblock In {\em 2023 IEEE Symposium on Security and Privacy (SP)}, pages 2676--2692. IEEE, 2023.

\bibitem{wu2025wildsync}
Wei-Cheng Wu, Stefan Nagy, and Christophe Hauser.
\newblock Wildsync: Automated fuzzing harness synthesis via wild api usage recovery.
\newblock volume~2, New York, NY, USA, June 2025. Association for Computing Machinery.

\bibitem{di2025hercules}
Luca Di~Bartolomeo, Philipp Mao, Yu-Jye Tung, Jessy Ayala, Samuele Doria, Paolo Celada, Marcel Busch, Joshua Garcia, Eleonora Losiouk, and Mathias Payer.
\newblock Hercules droidot and the murder on the jni express.
\newblock In {\em 34th USENIX Conference on Security Symposium}, SEC '25, USA, 2025. USENIX Association.

\bibitem{deng2023large}
Yinlin Deng, Chunqiu~Steven Xia, Chenyuan Yang, Shizhuo~Dylan Zhang, Shujing Yang, and Lingming Zhang.
\newblock Large language models are edge-case generators: Crafting unusual programs for fuzzing deep learning libraries.
\newblock In {\em 46th International Conference on Software Engineering}, ICSE '24, New York, NY, USA, 2024. Association for Computing Machinery.

\bibitem{xia2024fuzz4all}
Chunqiu~Steven Xia, Matteo Paltenghi, Jia Le~Tian, Michael Pradel, and Lingming Zhang.
\newblock Fuzz4all: Universal fuzzing with large language models.
\newblock In {\em 46th International Conference on Software Engineering}, pages 1--13, 2024.

\bibitem{yang2023whitefox}
Chenyuan Yang, Yinlin Deng, Runyu Lu, Jiayi Yao, Jiawei Liu, Reyhaneh Jabbarvand, and Lingming Zhang.
\newblock Whitefox: White-box compiler fuzzing empowered by large language models.
\newblock {\em Proceedings of the ACM on Programming Languages}, 8(OOPSLA2):709--735, 2024.

\bibitem{achiam2023gpt}
Josh Achiam, Steven Adler, Sandhini Agarwal, Lama Ahmad, Ilge Akkaya, Florencia~Leoni Aleman, Diogo Almeida, Janko Altenschmidt, Sam Altman, Shyamal Anadkat, et~al.
\newblock Gpt-4 technical report.
\newblock {\em arXiv preprint arXiv:2303.08774}, 2023.

\bibitem{li2023starcoder}
R~Li, LB~Allal, Y~Zi, N~Muennighoff, D~Kocetkov, C~Mou, M~Marone, C~Akiki, J~Li, J~Chim, et~al.
\newblock Starcoder: May the source be with you!
\newblock {\em Transactions on machine learning research}, 2023.

\bibitem{liu2024testing}
Zhe Liu, Chunyang Chen, Junjie Wang, Mengzhuo Chen, Boyu Wu, Zhilin Tian, Yuekai Huang, Jun Hu, and Qing Wang.
\newblock Testing the limits: Unusual text inputs generation for mobile app crash detection with large language model.
\newblock In {\em 46th International Conference on Software Engineering}, pages 1--12, 2024.

\bibitem{ELFUZZ}
Chuyang Chen, Brendan Dolan-Gavitt, and Zhiqiang Lin.
\newblock Elfuzz: efficient input generation via llm-driven synthesis over fuzzer space.
\newblock In {\em 34th USENIX Conference on Security Symposium}, USA, 2025. USENIX Association.

\bibitem{llmxps}
Yunpeng Tian, Feng Dong, Junhai Wang, Mu~Zhang, Zhiniang Peng, Zesen Ye, Xiapu Luo, and Haoyu Wang.
\newblock Error messages to fuzzing: Detecting xps parsing vulnerabilities in windows printing components.
\newblock In {\em 2025 ACM SIGSAC Conference on Computer and Communications Security}, CCS '25, page 798–812, New York, NY, USA, 2025. Association for Computing Machinery.

\bibitem{llmphp}
Yuancheng Jiang, Chuqi Zhang, Bonan Ruan, Jiahao Liu, Manuel Rigger, Roland H.~C. Yap, and Zhenkai Liang.
\newblock Fuzzing the php interpreter via dataflow fusion.
\newblock In {\em 34th USENIX Conference on Security Symposium}, SEC '25, USA, 2025. USENIX Association.

\bibitem{AIMFuzz}
Taewook Kim, Seokhyun Hong, and Yeongpil Cho.
\newblock Aimfuzz: Automated function-level in-memory fuzzing on binaries.
\newblock In {\em 19th ACM Asia Conference on Computer and Communications Security}, ASIA CCS '24, page 1510–1522, New York, NY, USA, 2024. Association for Computing Machinery.

\bibitem{sherman2025no}
Gabriel Sherman and Stefan Nagy.
\newblock No harness, no problem: Oracle-guided harnessing for auto-generating c api fuzzing harnesses.
\newblock In {\em 2025 IEEE/ACM 47th International Conference on Software Engineering (ICSE)}, pages 775--775. IEEE Computer Society, 2025.

\end{thebibliography}

\begin{appendices} 




\section*{Open Science}
\label{openscience}

To ensure transparency and reproducibility, we provide the following artifacts to facilitate the evaluation of our core contributions:

\noindent{\bf{Provided Artifacts:}}
\begin{itemize}
    \item An executable version of \tool (encrypted) along with experiment scripts and detailed documentation.
    \item Scripts to build and run the baseline tools (CKGFuzzer, OSS-Fuzz-Gen, PromeFuzz) for comparison.
    \item Harnesses generated by \tool for RQ1 (Branch Coverage, Section~\ref{subsec:rq1m}) and RQ2 (Bug Detection Ability, Section~\ref{subsec:rq2}).
    \item Complete steps and configurations to reproduce RQ4 (LLM Sensitivity, Section~\ref{subsec:rq4}) and RQ5 (Ablation Study, Section~\ref{subsec:rq2}).
\end{itemize}

\noindent{\bf{Artifact Limitations and Justifications:}}
\begin{itemize}
    \item The full source code of \tool will be released upon paper acceptance. For the review process, we provide an encrypted executable that runs in the specified environment and reproduces all experimental results.
    \item For RQ3, we \textbf{cannot} provide harnesses that trigger the CVE bugs due to responsible disclosure policies and anonymity. We will release these harnesses once the disclosure policies permit.
    \item Results from baseline tools are not included; we provide the build and execution scripts to ensure fair comparison can be performed.
\end{itemize}

\noindent{\bf{Anonymous Access for Reviewers:}}
\url{https://anonymous.4open.science/r/experiment-D67F}
(Read-only access, no credentials required)

\noindent{\bf{Target Projects:}}
All 25 evaluated projects are specified by their public GitHub URLs and commit hashes, allowing reviewers to build them independently.






\section*{Ethical Considerations}
This research focuses on automated harness generation, and we follow a responsible disclosure process. 
All identified vulnerabilities are promptly and privately reported to the responsible parties, with public disclosure only occurring after confirmation that the issues have been resolved. 
Since this study does not involve human subjects, no personally identifiable or sensitive data is included in the analyses or reports. 
For identified bugs, we adhere to established disclosure protocols, omitting undisclosed vulnerabilities and updating relevant sections progressively after official disclosure.

\section{Theoretical Proof of the Advantages of Decomposition and Rollback}
\label{appendix1}

Essentially, a fuzz harness refers to a piece of independent code responsible for feeding mutated input from fuzzers to targeted functions  accordingly. 
While current LLMs show substantial improvements in single-attempt code generation success rates (pass@1), their test harness code generation accuracy remains inadequate. 
To this issue, we decompose harness code generation into sequential stages mapping to improve precision. 
We mathematically demonstrate the staged decomposition and rollback's superiority in enhancing code generation accuracy through the following derivation.

The staged decomposition and rollback improves fault tolerance by:
\begin{itemize}
    \item decomposing complex tasks into discrete states with defined transition conditions;
    \item enabling rollback mechanisms at failed stages instead of terminating the entire process.
\end{itemize}

Let $P_o$ represents the success probability of the original harness generation task.
The task is decomposed into $n$ independent states where each state success probability $P_i$ satisfies $P_i > P_o^{1/n}$.

For sequential execution without retries, the compound success probability can be calculated as:

\begin{equation}
P_o^{\text{seq}} = \prod_{i=1}^{n} P_i
\end{equation}

In the staged rollback algorithm, allowing up to $k$ retries for each stage modifies the success probability of an individual stage to:

\begin{equation}
P_i' = 1 - (1 - P_i)^{k+1}
\end{equation}

When incorporating rollback with compensation success probability $c$, the modified system success probability is:

\begin{equation}
P_o^{\text{rollback}} = \prod_{i=1}^{n} \left(P_i' + (1 - P_i') \cdot c\right)
\end{equation}

The derivation proves $P_o^{\text{rollback}} > P_o^{\text{seq}} > P_o$, demonstrating the stage decomposition and rollback's dual advantage over both sequential execution and single-attempt approaches.

\section{Prompts Used in \tool}
\label{appendix4}

\subsection{Parameter Analysis}
\label{subsec:paraanal}
\begin{figure}[H]
	\centering
	\begin{tcolorbox}[colback=gray!5!white, colframe=gray!75!black, title=Parameter Analysis 1, fonttitle=\bfseries]
		\small
		\textbf{User Query:} ``Is the following \{parameter\_name\} a function parameter by reference?
The answer is output in json format. \\
Below is a sample output: 
\{
    "answer": "yes or no"
\}''

	\textbf{Expected Output:} ``\{"anser": "yes"\}'' or ``\{"anser": "no"\}''
	\end{tcolorbox}
	\label{p0_01}
\end{figure}



\begin{figure}[H]
	\centering
	\begin{tcolorbox}[colback=gray!5!white, colframe=gray!75!black, title=Parameter Analysis 2, fonttitle=\bfseries]
		\small
		\textbf{User Query:} ``Please list the following function reference parameters in detail.
The answer is output in json format.\\ 
Below is a sample output:
\{
    ``answer'': [param\_list]
\}''

	\textbf{Expected Output:} ``\{
    ``answer'': [param\_list]
\}''
	\end{tcolorbox}
	\label{p0_02}
\end{figure}



\begin{figure}[H]
	\centering
	\begin{tcolorbox}[colback=gray!5!white, colframe=gray!75!black, title=Parameter Analysis 3, fonttitle=\bfseries]
		\small
		\textbf{User Query:} ``Answer from the perspective of function input and output, what is the following situation of {parameter\_name} in the function?\\
A. Input \\B. Output \\C. Other\\
Below is a sample output: 
\{
    ``answer'': ``A''
\}''

	\textbf{Expected Output:} ``\{
    ``answer'': ``A''
\}''
	\end{tcolorbox}
	\label{p0_03}
\end{figure}



\subsection{Input Parameter Determination}
\label{subsec:inparadeter}
\begin{figure}[H]
	\centering
	\begin{tcolorbox}[colback=gray!5!white, colframe=gray!75!black, title=Input Parameter Determination 1, fonttitle=\bfseries]
		\small
		\textbf{User Query:} ``Identify the name of the input parameter in the following function that points to a parsed memory region, where this parameter points to a contiguous sequence of bytes or contiguous text data (excluding filenames or other semantically meaningful content), rather than pathnames, filenames, structs, floating-point numbers, or other specifically structured data. \\
Below is a sample output: 
\{
``pointer\_name:'' name of the parameter that points to the memory region, `none' if there is no such parameter"
\}''

	\textbf{Expected Output:} ``\{
``pointer\_name'': ``name of the parameter that points to the memory region, `none' if there is no such parameter.''
\}''
	\end{tcolorbox}
	\label{p1}
\end{figure}




\begin{figure}[H]
	\centering
	\begin{tcolorbox}[colback=gray!5!white, colframe=gray!75!black, title=Input Parameter Determination 2, fonttitle=\bfseries]
		\small
		\textbf{User Query:} ``Does the function parameter $Param$ below point to a contiguous byte stream? The region it points to should be a continuous sequence of bytes or contiguous text data (excluding filenames or other semantically meaningful content), rather than pathnames, filenames, structs, floating-point numbers, or other specifically structured data. \\
Below is a sample output:
\{ 
``answer'': ``yes'' or ``no'' 
\}''

	\textbf{Expected Output:} ``\{``answer'': ``yes''\}'' or ``\{``answer'': ``no''\}''
	\end{tcolorbox}
	\label{p2}
\end{figure}




\begin{figure}[H]
	\centering
	\begin{tcolorbox}[colback=gray!5!white, colframe=gray!75!black, title=Input Parameter Determination 3, fonttitle=\bfseries]
		\small
		\textbf{User Query:} ``Based on the following function, determine which of the following data types the parameter $Param$ belongs to:\\
A. Binary data\\
B. Text data\\
C. Struct\\
D. Floating-point number\\
E. Other \\
Below is a sample output:
\{
``answer'': ``A'' 
\}''

	\textbf{Expected Output:} ``\{``answer'': ``A''\}''
	\end{tcolorbox}
	\label{p2}
\end{figure}



\subsection{Functionality Analysis}
\label{subsec:funcanal}
\begin{figure}[H]
	\centering
	\begin{tcolorbox}[colback=gray!5!white, colframe=gray!75!black, title=Functionality Analysis 1, fonttitle=\bfseries]
		\small
		\textbf{User Query:} ``Which category does the following code belong to? Return in json format with reason: \\
A. Function processing \\
B. Function/resource initialization \\
C. Resource recovery \\
D. Test class \\
E. None of the above \\
Below is a sample output: 
\{
    ``reason'':``reason why select the answer'',
    ``answer'': ``A''
\}''

	\textbf{Expected Output:} ``\{``answer'': ``A''\}''
	\end{tcolorbox}
	\label{p0_1}
\end{figure}




\begin{figure}[H]
	\centering
	\begin{tcolorbox}[colback=gray!5!white, colframe=gray!75!black, title=Functionality Analysis 2, fonttitle=\bfseries]
		\small
		\textbf{User Query:} ``Is the following function a structure destruction function?
The answer is output in json format. 

Below is a sample output:
\{
``answer'': ``yes''
\}''

	\textbf{Expected Output:} ``\{``answer'': ``yes''\}'' or ``\{``answer'': ``no''\}''
	\end{tcolorbox}
	\label{p0_2}
\end{figure}




\subsection{Function Documentation Generation}
\label{subsec:funcdoc}
\begin{figure}[H]
	\centering
	\begin{tcolorbox}[colback=gray!5!white, colframe=gray!75!black, title=Function Documentation Generation, fonttitle=\bfseries]
		\small
		\textbf{User Query:} ``Analyze the function's functionality, and based on its usage scenario, provide a code snippet that calls the function in accordance with its scenario. The answer must strictly follow the markdown template below.
*Function Signature*
<!-- Briefly introduce the function's signature -->
*Functionality*
<!-- Briefly describe the function's functionality -->
*Application Scenario*
<!-- Briefly describe the function's usage scenario -->
*Example Code*
<!-- Provide an example code snippet that calls the function, ensuring the code is as concise as possible -->''

	\textbf{Expected Output:} ``- Function Signature:

...

- Functionality:

...

- Application Scenario:

...

- Example Code:

// example code here

''
	\end{tcolorbox}
	\label{p4-0}
\end{figure}


\subsection{Structure Snippet Stitching}
\label{subsec:stitch}
\begin{figure}[H]
	\centering
	\begin{tcolorbox}[colback=gray!5!white, colframe=gray!75!black, title=Structure Snippet Stitching, fonttitle=\bfseries]
		\small
		\textbf{User Query:} ``The following lists the functions and call snippets of \{len(func\_list)\} functions. 
According to the use scenarios of the functions and the logical relationship of the function processing the `{bytearray2str(one\_parameter.name)}' structure during the call process, write sample code that meets the call scenarios of \{len(func\_list)\} functions and \{func\_name\_str\} and calls \{len(func\_list)\} functions at the same time.
The answer should be as accurate and concise as possible. 
First, arrange the call sequence and the corresponding program structure according to the logical relationship of normal calls to these functions, output the reasons and then generate the code. 
The code should be optimized as much as possible and conform to the *logical call* relationship of the function functions.
Note: It must be ensured that \{len(func\_list)\} functions \{func\_name\_str\} must be called explicitly, and finally explain how \{func\_name\_str\} are called.
After the generation is completed, check the generated code to check whether the \{func\_name\_str\} functions are all called explicitly. 
If not, regenerate the code.
When generating, please note: 1. The functions and structures that appear in the code have been defined elsewhere, so there is no need to complete the code implementation. 2. The functions that appear in the code have been defined in the header file, so do not define them repeatedly. 3. Do not include functions that do not appear in the sample code.

The `\{parameter\_name\}' is as follows:
\{parameter\_body)\} \\
Below is a sample output:  
\{func\_api\_doc\}''

	\textbf{Expected Output:} `` // code snippet that invokes \{func\_list\} ''
	\end{tcolorbox}
	\label{p4}
\end{figure}

\subsection{Rough Code Assembly}
\label{subsec:codeass}
\begin{figure}[H]
	\centering
	\begin{tcolorbox}[colback=gray!5!white, colframe=gray!75!black, title=Rough Code Assembly, fonttitle=\bfseries]
		\small
		\textbf{User Query:} ``The following two code snippets with descriptions are the snippets for processing the structure \{parameter\_name\_list\_str\} and the structure \{parameter\_code.output\_all\_parameter\_name\_str()\} respectively. According to the logical relationship between the two code snippets and the function calls, give the code for processing the structure `\{parameter\_name\_list\_str\}` and the structure `\{bytearray2str(parameter\_code.parameter.name)\}` at the same time.
The answer should be as accurate and concise as possible. The functions appearing in the following code snippets should be called explicitly. The code should be optimized as much as possible and should conform to the logical calling relationship of the function functions.
Note: The generated code is all C language code, not C++. Do not invent code not mentioned above.\\
Below is a sample output:  
\{Content\}''

	\textbf{Expected Output:} `` // code that invokes entire FT in a main function ''
	\end{tcolorbox}
	\label{p5}
\end{figure}

\section{Target Details}
\label{app:rq1_detail}

\begin{table}[H]
  \centering
  \caption{Detailed project information: commit hash and star counts.}
    \begin{tabular}{c|c|c}
    \toprule
    \textbf{Project} & \textbf{Commit ID (Version)} & \textbf{Stars} \\
    \midrule
        \rowcolor[HTML]{EFEFEF}
    c-ares & 16c873c & 2k \\
    fribidi & b28f43b & 0.4k \\
        \rowcolor[HTML]{EFEFEF}
    libyaml & 840b65c & 1k \\
    cjson & 12c4bf1 & 11.5k \\
        \rowcolor[HTML]{EFEFEF}
    sqlite3 & 3.51.1 & - \\
    zlib  & 5a82f71 & 6.1k \\
        \rowcolor[HTML]{EFEFEF}
    opus  & 2785f8d & 2.6k \\
    libxml2 & 1039cd5 & 0.6k \\
        \rowcolor[HTML]{EFEFEF}
    lz4   & 67a385a & 10.9k \\
    libssh2 & 8a871d0 & 1.4k \\
        \rowcolor[HTML]{EFEFEF}
    expat & 5dd4a63 & 1.2k \\
    avahi & 6d801ab & 1.3k \\
        \rowcolor[HTML]{EFEFEF}
    libxslt & c8b1ea4 & 0.1k \\
    libtiff & d31dd34 & - \\
        \rowcolor[HTML]{EFEFEF}
    lcms  & 9ac94ea & 0.6k \\
    libarchive & dcbf1e0 & 3.4k \\
        \rowcolor[HTML]{EFEFEF}
    xz    & dd4a1b2 & 0.9k \\
    jq    & 0b1ef46 & 33.2k \\
        \rowcolor[HTML]{EFEFEF}
    pjsip & 6c24457 & 2.5k \\
    kamailio & a05a758 & 2.7k \\
        \rowcolor[HTML]{EFEFEF}
    postfix & e66967d & 0.5k \\
    gdbm  & 0d3d46c & - \\
        \rowcolor[HTML]{EFEFEF}
    file  & 5184ca2 & 1.4k \\
    hpn-ssh & c03bd45 & 0.5k \\
        \rowcolor[HTML]{EFEFEF}
    dropbear & bd12a86 & 1.9k \\
    \bottomrule
    \end{tabular}%
  \label{project_detail}%
\end{table}%

\section{Operations of collecting branch coverage for all baselines.}
\label{Collect_branch_op}

\begin{lstlisting}[caption={Operations of collecting branch coverage.}, label=hallucination_case]
# copy all profile(*.profraw) and fuzzer(*.out) to workdir
> cd workdir
> llvm-profdata merge -sparse *.profraw -o merged.profdata
> for fuzzer in "*.out"; do
    llvm-cov report ${fuzzer} -instr-profile=merged.profdata > ${fuzzer}_cov_report.txt
>done
> llvm-cov show fuzzer_with_max_total_branches.out -instr-profile=merged.profdata --format=html -output-dir=final_coverage
\end{lstlisting}

\section{Efficiency Statistics Details}
\label{app:eff_details}

\begin{table*}[htbp]
  \centering
  \caption{Detail results of efficiency. PreT: pre-processing time (hours), GenT: generation time per harness (hours), Tokens: token consumption (millions), \#Harness: number of harness generated. OF: OSS-Fuzz-Gen, CK: CKGFuzzer, PF: PromeFuzz, SF: \tool. \#S. denotes the average number of stages executed per successfully generated harness for \tool.}
    \begin{tabular}{c|cccc|cccc|cccc|cccc|c}
    \toprule
    \multirow{2}[4]{*}{\textbf{Project}} & \multicolumn{4}{c|}{\textbf{PreT (h)}}  & \multicolumn{4}{c|}{\textbf{GenT (h)}}  & \multicolumn{4}{c|}{\textbf{Tokens (M)}} & \multicolumn{4}{c|}{\textbf{\#Harness}} & \multirow{2}[4]{*}{\textbf{\#S.}} \\
    \cmidrule(lr){2-5} \cmidrule(lr){6-9} \cmidrule(lr){10-13} \cmidrule(lr){14-17}
        & \textbf{OF} & \textbf{CK} & \textbf{PF} & \textbf{SF} & \textbf{OF} & \textbf{CK} & \textbf{PF} & \textbf{SF} & \textbf{OF} & \textbf{CK} & \textbf{PF} & \textbf{SF} & \textbf{OF} & \textbf{CK} & \textbf{PF} & \textbf{SF} & \\
    \midrule
        \rowcolor[HTML]{EFEFEF}
    c-ares    & - & 4.02  & 3.32  & 2.23  & 0.42  & 0.25  & 0.17  & 0.12  & 20.25  & 8.72  & 4.02  & 3.47  & 74    & 100   & 94    & 164   & 7.22  \\
    fribidi   & -  & 3.97  & 0.52  & 0.25  & 0.13  & 0.86  & 0.22  & 0.22  & 2.31  & 5.78  & 1.02  & 1.86  & 32    & 15    & 16    & 30    & 7.72  \\
        \rowcolor[HTML]{EFEFEF}
    libyaml   & -  & 4.07  & 0.30  & 0.37  & 0.20  & 0.66  & 0.10  & 0.23  & 3.19  & 5.31  & 0.85  & 2.13  & 27    & 30    & 28    & 36    & 7.46  \\
    cjson     & -  & 3.83  & 2.67  & 0.42  & 0.11  & 0.43  & 0.08  & 0.21  & 3.72  & 6.77  & 1.78  & 7.75  & 58    & 40    & 64    & 150   & 7.08  \\
        \rowcolor[HTML]{EFEFEF}
    sqlite3   & -  & 5.47  & 8.62  & 4.48  & 1.46  & 1.18  & 1.17  & 0.28  & 44.89  & 38.09  & 37.87 & 49.15 & 45    & 112   & 104   & 603   & 7.33  \\
    zlib      & - & 3.93  & 1.55  & 1.33  & 0.17  & 0.40  & 0.10  & 0.15  & 10.78  & 9.16  & 1.79  & 4.77  & 104   & 69    & 63    & 156   & 7.04  \\
        \rowcolor[HTML]{EFEFEF}
    opus      & -  & 4.63  & 7.32  & 2.68  & 6.23  & 3.19  & 2.84  & 0.11  & 24.49  & 28.16  & 32.22 & 9.01  & 6     & 24    & 29    & 377   & 7.28  \\
    libxml2   & - & 4.35  & 5.15  & 5.03  & 0.32  & 0.43  & 0.84  & 0.18  & 30.22  & 17.32  & 19.53 & 16.13 & 147   & 97    & 53    & 266   & 7.29  \\
        \rowcolor[HTML]{EFEFEF}
    lz4       &-  & 3.85  & 0.72  & 1.10  & 0.26  & 0.91  & 0.30  & 0.16  & 11.21  & 9.54  & 2.75  & 3.74  & 65    & 42    & 44    & 158   & 7.22  \\
    libssh2   & -  & 4.52  & 3.88  & 1.88  & 0.27  & 1.15  & 1.56  & 0.26  & 18.92  & 10.37  & 11.74 & 5.15  & 105   & 21    & 17    & 55    & 7.28  \\
        \rowcolor[HTML]{EFEFEF}
    expat     &- & 4.27  & 3.05  & 1.30  & 0.55  & 0.79  & 0.47  & 0.36  & 9.77  & 9.95  & 6.44  & 9.52  & 30    & 50    & 58    & 134   & 6.98  \\
    avahi     &-  & 4.45  & 3.77  & 3.03  & 0.74  & 1.05  & 0.39  & 0.20  & 22.23  & 18.69  & 15.83 & 9.27  & 50    & 36    & 84    & 126   & 7.49  \\
        \rowcolor[HTML]{EFEFEF}
    libxslt   & - & 3.87  & 2.85  & 5.23  & 0.18  & 0.69  & 1.19  & 0.61  & 15.29  & 19.48  & 21.7  & 16.64 & 153   & 74    & 62    & 121   & 7.24  \\
    libtiff   & -  & 4.87  & 4.92  & 2.10  & 0.40  & 0.45  & 0.20  & 0.17  & 42.85  & 23.00  & 7.26  & 13.19 & 128   & 153   & 128   & 353   & 11.14 \\
        \rowcolor[HTML]{EFEFEF}
    lcms      & - & 4.15  & 3.57  & 2.27  & 0.19  & 0.22  & 0.20  & 0.26  & 26.48  & 24.49  & 13.69 & 27.02 & 200   & 234   & 141   & 245   & 8.31  \\
    libarchive& - & 4.93  & 10.17 & 2.65  & 0.70  & 2.06  & 1.57  & 0.24  & 35.52  & 41.07  & 54.75 & 43.38 & 68    & 76    & 124   & 709   & 8.00  \\
        \rowcolor[HTML]{EFEFEF}
    xz        & -  & 4.48  & 10.40 & 1.50  & 0.35  & 1.43  & 0.93  & 0.27  & 11.88  & 22.34  & 20.41 & 12.7  & 55    & 46    & 60    & 157   & 7.17  \\
    jq        & -  & N/A   & 3.25  & 2.05  & 2.10  & N/A   & 2.88  & 0.35  & 17.24  & N/A   & 16.85 & 8.39  & 13    & N/A   & 10    & 57    & 7.09  \\
        \rowcolor[HTML]{EFEFEF}
    pjsip     & -  & N/A   & 11.77 & 10.82 & 14.28 & N/A   & 4.61  & 0.46  & 123.11 & N/A   & 142.95& 45.3  & 12    & N/A   & 75    & 405   & 8.11  \\
    kamailio  & -  & N/A   & 16.17 & 28.85 & 2.49  & N/A   & 2.70  & 0.63  & 140.58 & N/A   & 76.85 & 53.32 & 43    & N/A   & 87    & 316   & 9.12  \\
        \rowcolor[HTML]{EFEFEF}
    postfix   & -  & N/A   & 10.92 & 5.02  & 0.89  & N/A   & 6.32  & 0.21  & 90.89  & N/A   & 77.59 & 36.35 & 125   & N/A   & 38    & 959   & 8.60  \\
    gdbm      & -  & N/A   & 12.82 & 0.72  & 0.53  & N/A   & 1.40  & 0.20  & 6.82   & N/A   & 16.82 & 7.18  & 20    & N/A   & 27    & 113   & 6.92  \\
        \rowcolor[HTML]{EFEFEF}
    file      & - & N/A   & 10.60 & 1.55  & 0.09  & N/A   & 0.10  & 0.50  & 18.47  & N/A   & 2.37  & 14.12 & 290   & N/A   & 90    & 162   & 7.99  \\
    hpn-ssh   & -  & N/A   & 18.28 & 6.07  & 6.85  & N/A   & 3.17  & 0.85  & 120.79 & N/A   & 174.96& 161.91& 21    & N/A   & 122   & 474   & 9.02  \\
        \rowcolor[HTML]{EFEFEF}
    dropbear  & -  & N/A   & 9.47  & 2.98  & 1.35  & N/A   & 1.45  & 1.12  & 34.78  & N/A   & 19.35 & 27.22 & 36    & N/A   & 45    & 85    & 8.28  \\
    \bottomrule
    \end{tabular}%
  \label{tab:appendix_efficiency}%
\end{table*}

\end{appendices}

\end{document}